\documentclass[11pt]{article}
\pdfoutput=1

\usepackage{jcappub}

\usepackage{color, xcolor, appendix,latexsym,amsmath,amssymb,graphicx,booktabs,epsfig,hyperref,url,nicefrac,diagbox,multirow}
\usepackage{multicol}
\usepackage{tablefootnote}
\usepackage{bold-extra}

\usepackage{bm} 

\numberwithin{equation}{section}
\usepackage[english]{babel}
\usepackage{subcaption}
\newcommand{\msun}{M$_\odot$}
\usepackage{multirow}
\usepackage{wrapfig}
\usepackage{lineno}
\usepackage[paper=a4paper,tmargin=3.5cm,bmargin=1.5cm,lmargin=2cm,rmargin=2cm]{geometry}

\definecolor{MyBlue}{rgb}{0.15,0.15,0.70}
\hypersetup{
colorlinks=true,
citecolor=MyBlue,
linkcolor=MyBlue,
urlcolor=MyBlue
}
\definecolor{lightgray}{gray}{0.9}

\newcommand{\princess}{{\sc Princess}}

\usepackage{orcidlink}

    \usepackage[breakable]{tcolorbox}
    \usepackage{parskip} 

    \usepackage{textcomp} 
    \AtBeginDocument{%
        \def\PYZsq{\textquotesingle}
    }
    \usepackage{upquote} 
    \usepackage{eurosym} 
    \usepackage{dirtree}
    \usepackage[mathletters]{ucs} 
    \usepackage{fancyvrb} 
    \usepackage{grffile} 
    \makeatletter 
    \@ifpackagelater{grffile}{2019/11/01}
    {
    }
    {
      \def\Gread@@xetex#1{%
        \IfFileExists{"\Gin@base".bb}%
        {\Gread@eps{\Gin@base.bb}}%
        {\Gread@@xetex@aux#1}%
      }
    }
    \makeatother
    \usepackage[Export]{adjustbox} 
    \adjustboxset{max size={0.9\linewidth}{0.9\paperheight}}

    \usepackage{hyperref}
    \usepackage{longtable} 
    \usepackage{booktabs}  
    \usepackage[inline]{enumitem} 
    \usepackage[normalem]{ulem} 
    \usepackage{mathrsfs}

    \definecolor{urlcolor}{rgb}{0,.145,.698}
    \definecolor{linkcolor}{rgb}{.71,0.21,0.01}
    \definecolor{citecolor}{rgb}{.12,.54,.11}

    \definecolor{ansi-black}{HTML}{3E424D}
    \definecolor{ansi-black-intense}{HTML}{282C36}
    \definecolor{ansi-red}{HTML}{E75C58}
    \definecolor{ansi-red-intense}{HTML}{B22B31}
    \definecolor{ansi-green}{HTML}{00A250}
    \definecolor{ansi-green-intense}{HTML}{007427}
    \definecolor{ansi-yellow}{HTML}{DDB62B}
    \definecolor{ansi-yellow-intense}{HTML}{B27D12}
    \definecolor{ansi-blue}{HTML}{208FFB}
    \definecolor{ansi-blue-intense}{HTML}{0065CA}
    \definecolor{ansi-magenta}{HTML}{D160C4}
    \definecolor{ansi-magenta-intense}{HTML}{A03196}
    \definecolor{ansi-cyan}{HTML}{60C6C8}
    \definecolor{ansi-cyan-intense}{HTML}{258F8F}
    \definecolor{ansi-white}{HTML}{C5C1B4}
    \definecolor{ansi-white-intense}{HTML}{A1A6B2}
    \definecolor{ansi-default-inverse-fg}{HTML}{FFFFFF}
    \definecolor{ansi-default-inverse-bg}{HTML}{000000}

    \definecolor{outerrorbackground}{HTML}{FFDFDF}

    \DefineVerbatimEnvironment{Highlighting}{Verbatim}{commandchars=\\\{\}}

    \let\Oldtex\TeX
    \let\Oldlatex\LaTeX
    \renewcommand{\TeX}{\textrm{\Oldtex}}
    \renewcommand{\LaTeX}{\textrm{\Oldlatex}}

\makeatletter
\def\PY@reset{\let\PY@it=\relax \let\PY@bf=\relax%
    \let\PY@ul=\relax \let\PY@tc=\relax%
    \let\PY@bc=\relax \let\PY@ff=\relax}
\def\PY@tok#1{\csname PY@tok@#1\endcsname}
\def\PY@toks#1+{\ifx\relax#1\empty\else%
    \PY@tok{#1}\expandafter\PY@toks\fi}
\def\PY@do#1{\PY@bc{\PY@tc{\PY@ul{%
    \PY@it{\PY@bf{\PY@ff{#1}}}}}}}
\def\PY#1#2{\PY@reset\PY@toks#1+\relax+\PY@do{#2}}

\@namedef{PY@tok@w}{\def\PY@tc##1{\textcolor[rgb]{0.73,0.73,0.73}{##1}}}
\@namedef{PY@tok@c}{\let\PY@it=\textit\def\PY@tc##1{\textcolor[rgb]{0.25,0.50,0.50}{##1}}}
\@namedef{PY@tok@cp}{\def\PY@tc##1{\textcolor[rgb]{0.74,0.48,0.00}{##1}}}
\@namedef{PY@tok@k}{\let\PY@bf=\textbf\def\PY@tc##1{\textcolor[rgb]{0.00,0.50,0.00}{##1}}}
\@namedef{PY@tok@kp}{\def\PY@tc##1{\textcolor[rgb]{0.00,0.50,0.00}{##1}}}
\@namedef{PY@tok@kt}{\def\PY@tc##1{\textcolor[rgb]{0.69,0.00,0.25}{##1}}}
\@namedef{PY@tok@o}{\def\PY@tc##1{\textcolor[rgb]{0.40,0.40,0.40}{##1}}}
\@namedef{PY@tok@ow}{\let\PY@bf=\textbf\def\PY@tc##1{\textcolor[rgb]{0.67,0.13,1.00}{##1}}}
\@namedef{PY@tok@nb}{\def\PY@tc##1{\textcolor[rgb]{0.00,0.50,0.00}{##1}}}
\@namedef{PY@tok@nf}{\def\PY@tc##1{\textcolor[rgb]{0.00,0.00,1.00}{##1}}}
\@namedef{PY@tok@nc}{\let\PY@bf=\textbf\def\PY@tc##1{\textcolor[rgb]{0.00,0.00,1.00}{##1}}}
\@namedef{PY@tok@nn}{\let\PY@bf=\textbf\def\PY@tc##1{\textcolor[rgb]{0.00,0.00,1.00}{##1}}}
\@namedef{PY@tok@ne}{\let\PY@bf=\textbf\def\PY@tc##1{\textcolor[rgb]{0.82,0.25,0.23}{##1}}}
\@namedef{PY@tok@nv}{\def\PY@tc##1{\textcolor[rgb]{0.10,0.09,0.49}{##1}}}
\@namedef{PY@tok@no}{\def\PY@tc##1{\textcolor[rgb]{0.53,0.00,0.00}{##1}}}
\@namedef{PY@tok@nl}{\def\PY@tc##1{\textcolor[rgb]{0.63,0.63,0.00}{##1}}}
\@namedef{PY@tok@ni}{\let\PY@bf=\textbf\def\PY@tc##1{\textcolor[rgb]{0.60,0.60,0.60}{##1}}}
\@namedef{PY@tok@na}{\def\PY@tc##1{\textcolor[rgb]{0.49,0.56,0.16}{##1}}}
\@namedef{PY@tok@nt}{\let\PY@bf=\textbf\def\PY@tc##1{\textcolor[rgb]{0.00,0.50,0.00}{##1}}}
\@namedef{PY@tok@nd}{\def\PY@tc##1{\textcolor[rgb]{0.67,0.13,1.00}{##1}}}
\@namedef{PY@tok@s}{\def\PY@tc##1{\textcolor[rgb]{0.73,0.13,0.13}{##1}}}
\@namedef{PY@tok@sd}{\let\PY@it=\textit\def\PY@tc##1{\textcolor[rgb]{0.73,0.13,0.13}{##1}}}
\@namedef{PY@tok@si}{\let\PY@bf=\textbf\def\PY@tc##1{\textcolor[rgb]{0.73,0.40,0.53}{##1}}}
\@namedef{PY@tok@se}{\let\PY@bf=\textbf\def\PY@tc##1{\textcolor[rgb]{0.73,0.40,0.13}{##1}}}
\@namedef{PY@tok@sr}{\def\PY@tc##1{\textcolor[rgb]{0.73,0.40,0.53}{##1}}}
\@namedef{PY@tok@ss}{\def\PY@tc##1{\textcolor[rgb]{0.10,0.09,0.49}{##1}}}
\@namedef{PY@tok@sx}{\def\PY@tc##1{\textcolor[rgb]{0.00,0.50,0.00}{##1}}}
\@namedef{PY@tok@m}{\def\PY@tc##1{\textcolor[rgb]{0.40,0.40,0.40}{##1}}}
\@namedef{PY@tok@gh}{\let\PY@bf=\textbf\def\PY@tc##1{\textcolor[rgb]{0.00,0.00,0.50}{##1}}}
\@namedef{PY@tok@gu}{\let\PY@bf=\textbf\def\PY@tc##1{\textcolor[rgb]{0.50,0.00,0.50}{##1}}}
\@namedef{PY@tok@gd}{\def\PY@tc##1{\textcolor[rgb]{0.63,0.00,0.00}{##1}}}
\@namedef{PY@tok@gi}{\def\PY@tc##1{\textcolor[rgb]{0.00,0.63,0.00}{##1}}}
\@namedef{PY@tok@gr}{\def\PY@tc##1{\textcolor[rgb]{1.00,0.00,0.00}{##1}}}
\@namedef{PY@tok@ge}{\let\PY@it=\textit}
\@namedef{PY@tok@gs}{\let\PY@bf=\textbf}
\@namedef{PY@tok@gp}{\let\PY@bf=\textbf\def\PY@tc##1{\textcolor[rgb]{0.00,0.00,0.50}{##1}}}
\@namedef{PY@tok@go}{\def\PY@tc##1{\textcolor[rgb]{0.53,0.53,0.53}{##1}}}
\@namedef{PY@tok@gt}{\def\PY@tc##1{\textcolor[rgb]{0.00,0.27,0.87}{##1}}}
\@namedef{PY@tok@err}{\def\PY@bc##1{{\setlength{\fboxsep}{\string -\fboxrule}\fcolorbox[rgb]{1.00,0.00,0.00}{1,1,1}{\strut ##1}}}}
\@namedef{PY@tok@kc}{\let\PY@bf=\textbf\def\PY@tc##1{\textcolor[rgb]{0.00,0.50,0.00}{##1}}}
\@namedef{PY@tok@kd}{\let\PY@bf=\textbf\def\PY@tc##1{\textcolor[rgb]{0.00,0.50,0.00}{##1}}}
\@namedef{PY@tok@kn}{\let\PY@bf=\textbf\def\PY@tc##1{\textcolor[rgb]{0.00,0.50,0.00}{##1}}}
\@namedef{PY@tok@kr}{\let\PY@bf=\textbf\def\PY@tc##1{\textcolor[rgb]{0.00,0.50,0.00}{##1}}}
\@namedef{PY@tok@bp}{\def\PY@tc##1{\textcolor[rgb]{0.00,0.50,0.00}{##1}}}
\@namedef{PY@tok@fm}{\def\PY@tc##1{\textcolor[rgb]{0.00,0.00,1.00}{##1}}}
\@namedef{PY@tok@vc}{\def\PY@tc##1{\textcolor[rgb]{0.10,0.09,0.49}{##1}}}
\@namedef{PY@tok@vg}{\def\PY@tc##1{\textcolor[rgb]{0.10,0.09,0.49}{##1}}}
\@namedef{PY@tok@vi}{\def\PY@tc##1{\textcolor[rgb]{0.10,0.09,0.49}{##1}}}
\@namedef{PY@tok@vm}{\def\PY@tc##1{\textcolor[rgb]{0.10,0.09,0.49}{##1}}}
\@namedef{PY@tok@sa}{\def\PY@tc##1{\textcolor[rgb]{0.73,0.13,0.13}{##1}}}
\@namedef{PY@tok@sb}{\def\PY@tc##1{\textcolor[rgb]{0.73,0.13,0.13}{##1}}}
\@namedef{PY@tok@sc}{\def\PY@tc##1{\textcolor[rgb]{0.73,0.13,0.13}{##1}}}
\@namedef{PY@tok@dl}{\def\PY@tc##1{\textcolor[rgb]{0.73,0.13,0.13}{##1}}}
\@namedef{PY@tok@s2}{\def\PY@tc##1{\textcolor[rgb]{0.73,0.13,0.13}{##1}}}
\@namedef{PY@tok@sh}{\def\PY@tc##1{\textcolor[rgb]{0.73,0.13,0.13}{##1}}}
\@namedef{PY@tok@s1}{\def\PY@tc##1{\textcolor[rgb]{0.73,0.13,0.13}{##1}}}
\@namedef{PY@tok@mb}{\def\PY@tc##1{\textcolor[rgb]{0.40,0.40,0.40}{##1}}}
\@namedef{PY@tok@mf}{\def\PY@tc##1{\textcolor[rgb]{0.40,0.40,0.40}{##1}}}
\@namedef{PY@tok@mh}{\def\PY@tc##1{\textcolor[rgb]{0.40,0.40,0.40}{##1}}}
\@namedef{PY@tok@mi}{\def\PY@tc##1{\textcolor[rgb]{0.40,0.40,0.40}{##1}}}
\@namedef{PY@tok@il}{\def\PY@tc##1{\textcolor[rgb]{0.40,0.40,0.40}{##1}}}
\@namedef{PY@tok@mo}{\def\PY@tc##1{\textcolor[rgb]{0.40,0.40,0.40}{##1}}}
\@namedef{PY@tok@ch}{\let\PY@it=\textit\def\PY@tc##1{\textcolor[rgb]{0.25,0.50,0.50}{##1}}}
\@namedef{PY@tok@cm}{\let\PY@it=\textit\def\PY@tc##1{\textcolor[rgb]{0.25,0.50,0.50}{##1}}}
\@namedef{PY@tok@cpf}{\let\PY@it=\textit\def\PY@tc##1{\textcolor[rgb]{0.25,0.50,0.50}{##1}}}
\@namedef{PY@tok@c1}{\let\PY@it=\textit\def\PY@tc##1{\textcolor[rgb]{0.25,0.50,0.50}{##1}}}
\@namedef{PY@tok@cs}{\let\PY@it=\textit\def\PY@tc##1{\textcolor[rgb]{0.25,0.50,0.50}{##1}}}

\def\PYZus{\char`\_}
\def\PYZob{\char`\{}
\def\PYZcb{\char`\}}

\def\PYZsh{\char`\#}

\def\PYZhy{\char`\-}
\def\PYZsq{\char`\'}
\def\PYZdq{\char`\"}

\makeatother

    \makeatletter
        \newbox\Wrappedcontinuationbox 
        \newbox\Wrappedvisiblespacebox 
        \newcommand*\Wrappedvisiblespace {\textcolor{red}{\textvisiblespace}} 
        \newcommand*\Wrappedcontinuationsymbol {\textcolor{red}{\llap{\tiny$\m@th\hookrightarrow$}}} 
        \newcommand*\Wrappedcontinuationindent {3ex } 
        \newcommand*\Wrappedafterbreak {\kern\Wrappedcontinuationindent\copy\Wrappedcontinuationbox} 
        \newcommand*\Wrappedbreaksatspecials {%
            \def\PYGZus{\discretionary{\char`\_}{\Wrappedafterbreak}{\char`\_}}%
            \def\PYGZob{\discretionary{}{\Wrappedafterbreak\char`\{}{\char`\{}}%
            \def\PYGZcb{\discretionary{\char`\}}{\Wrappedafterbreak}{\char`\}}}%
            \def\PYGZca{\discretionary{\char`\^}{\Wrappedafterbreak}{\char`\^}}%
            \def\PYGZam{\discretionary{\char`\&}{\Wrappedafterbreak}{\char`\&}}%
            \def\PYGZlt{\discretionary{}{\Wrappedafterbreak\char`\<}{\char`\<}}%
            \def\PYGZgt{\discretionary{\char`\>}{\Wrappedafterbreak}{\char`\>}}%
            \def\PYGZsh{\discretionary{}{\Wrappedafterbreak\char`\#}{\char`\#}}%
            \def\PYGZpc{\discretionary{}{\Wrappedafterbreak\char`\%}{\char`\%}}%
            \def\PYGZdl{\discretionary{}{\Wrappedafterbreak\char`\$}{\char`\$}}%
            \def\PYGZhy{\discretionary{\char`\-}{\Wrappedafterbreak}{\char`\-}}%
            \def\PYGZsq{\discretionary{}{\Wrappedafterbreak\textquotesingle}{\textquotesingle}}%
            \def\PYGZdq{\discretionary{}{\Wrappedafterbreak\char`\"}{\char`\"}}%
            \def\PYGZti{\discretionary{\char`\~}{\Wrappedafterbreak}{\char`\~}}%
        } 
        \newcommand*\Wrappedbreaksatpunct {%
            \lccode`\~`\.\lowercase{\def~}{\discretionary{\hbox{\char`\.}}{\Wrappedafterbreak}{\hbox{\char`\.}}}%
            \lccode`\~`\,\lowercase{\def~}{\discretionary{\hbox{\char`\,}}{\Wrappedafterbreak}{\hbox{\char`\,}}}%
            \lccode`\~`\;\lowercase{\def~}{\discretionary{\hbox{\char`\;}}{\Wrappedafterbreak}{\hbox{\char`\;}}}%
            \lccode`\~`\:\lowercase{\def~}{\discretionary{\hbox{\char`\:}}{\Wrappedafterbreak}{\hbox{\char`\:}}}%
            \lccode`\~`\?\lowercase{\def~}{\discretionary{\hbox{\char`\?}}{\Wrappedafterbreak}{\hbox{\char`\?}}}%
            \lccode`\~`\!\lowercase{\def~}{\discretionary{\hbox{\char`\!}}{\Wrappedafterbreak}{\hbox{\char`\!}}}%
            \lccode`\~`\/\lowercase{\def~}{\discretionary{\hbox{\char`\/}}{\Wrappedafterbreak}{\hbox{\char`\/}}}%
            \catcode`\.\active
            \catcode`\,\active 
            \catcode`\;\active
            \catcode`\:\active
            \catcode`\?\active
            \catcode`\!\active
            \catcode`\/\active 
            \lccode`\~`\~ 	
        }
    \makeatother

    \let\OriginalVerbatim=\Verbatim
    \makeatletter
    \renewcommand{\Verbatim}[1][1]{%
        \sbox\Wrappedcontinuationbox {\Wrappedcontinuationsymbol}%
        \sbox\Wrappedvisiblespacebox {\FV@SetupFont\Wrappedvisiblespace}%
        \def\FancyVerbFormatLine ##1{\hsize\linewidth
            \vtop{\raggedright\hyphenpenalty\z@\exhyphenpenalty\z@
                \doublehyphendemerits\z@\finalhyphendemerits\z@
                \strut ##1\strut}%
        }%
        \def\FV@Space {%
            \nobreak\hskip\z@ plus\fontdimen3\font minus\fontdimen4\font
            \discretionary{\copy\Wrappedvisiblespacebox}{\Wrappedafterbreak}
            {\kern\fontdimen2\font}%
        }%
        
        \Wrappedbreaksatspecials
        \OriginalVerbatim[#1,codes*=\Wrappedbreaksatpunct]%
    }
    \makeatother

    \definecolor{incolor}{HTML}{303F9F}
    \definecolor{outcolor}{HTML}{D84315}
    \definecolor{cellborder}{HTML}{CFCFCF}
    \definecolor{cellbackground}{HTML}{F7F7F7}
    
    \makeatletter
    \newcommand{\boxspacing}{\kern\kvtcb@left@rule\kern\kvtcb@boxsep}
    \makeatother
    \newcommand{\prompt}[4]{
        {\ttfamily\llap{{\color{#2}[#3]:\hspace{3pt}#4}}\vspace{-\baselineskip}}
    }
    
\title{PRINCESS: \\ Prediction of compact binaries observations with gravitational waves \\
{\small Version 1.0, January 2025}}

 \author[a,b,c,\orcidlink{0000-0002-9779-2838}]{Carole~P\'{e}rigois,\note{Corresponding author.}}

\affiliation[a]{ Physics and Astronomy Department Galileo Galilei, University of Padova,\\Vicolo dell’Osservatorio 3, I–35122, Padova, Italy}
\affiliation[b]{ INFN - Padova,\\Via Marzolo 8, I–35131 Padova, Italy}
\affiliation[c]{ INAF - Osservatorio Astronomico di Padova,\\Vicolo dell’Osservatorio 5, I-35122 Padova, Italy}

\emailAdd{caroleperigois@outlook.com}

\abstract{
We present \princess, a computational tool designed to predict gravitational wave observations from compact binary coalescences (CBCs) in current and future detector networks. PRINCESS uniquely combines predictions of both individual gravitational wave events and the associated astrophysical background, leveraging user-provided CBC catalogs. With the anticipated improvements in detector sensitivity from second-generation (2G) to third-generation (3G) observatories like the Einstein Telescope and Cosmic Explorer, the tool aims to constrain models of stellar formation and compact object evolution. PRINCESS calculates the signal-to-noise ratio (SNR) for individual events and predicts the stochastic background arising from unresolved sources. We detail the code’s structure, installation, and usage, providing examples of predictions for different astrophysical models and detector configurations. Results include forecasts for binary black hole detections and background spectra, highlighting the potential of future networks to resolve nearly all CBC events. We also discuss ongoing developments to expand PRINCESS's capabilities, including accounting for additional sources such as extreme mass ratio inspirals (EMRIs) and incorporating more advanced detector models.
}

\begin{document}

\begin{wrapfigure}{r}{0.30\textwidth}
\vspace{+2.5cm}
\includegraphics[width=0.28\textwidth]{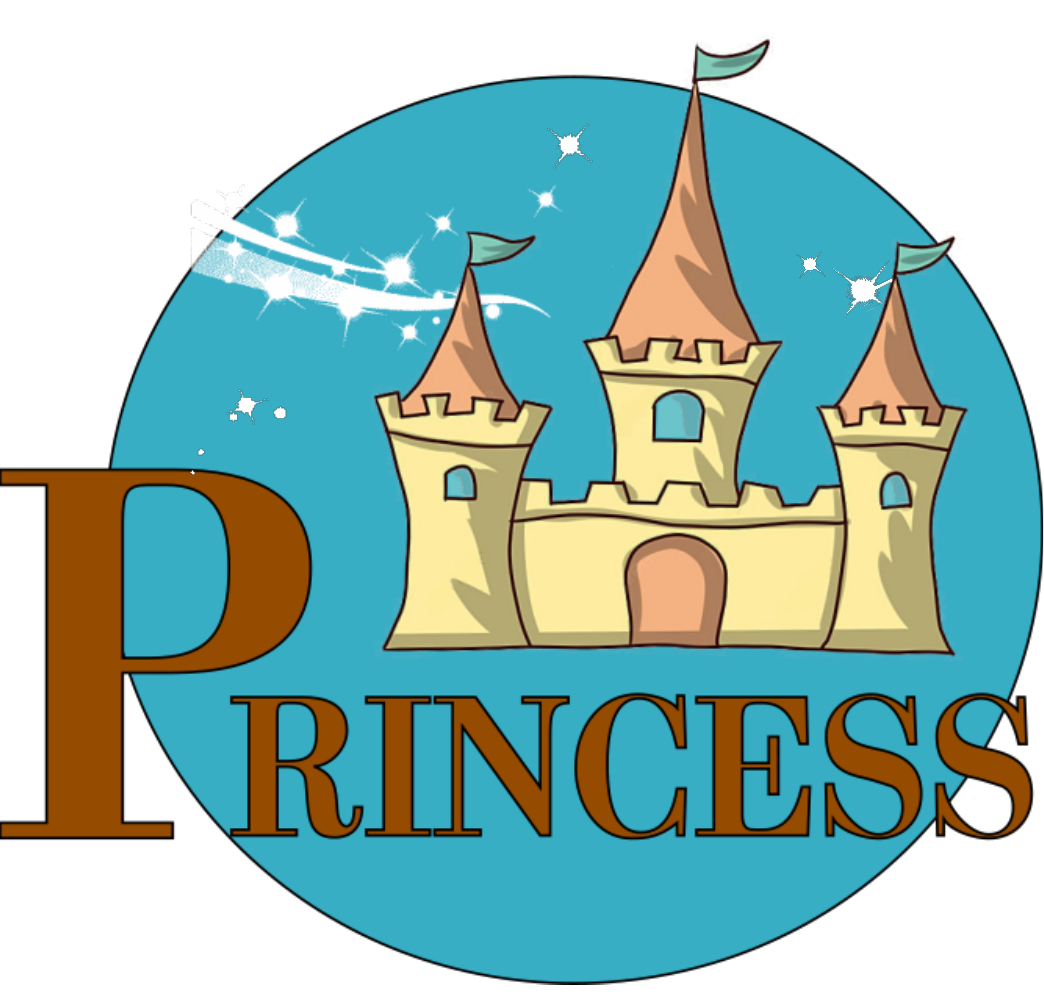} 
\end{wrapfigure}
\maketitle

\section{Introduction}

The first direct detection of gravitational waves (GWs) by the LIGO\cite{LIGO_AdLIGO_2015} and Virgo\cite{Virgo_adVirgo_2015} collaborations in 2015, from the merger of two stellar-mass black holes (GW150914, \cite{LIGO_2016a, LIGO_GW150914}), not only confirmed Einstein’s General Theory of Relativity \cite{LIGO_2016_GR} but also opened a new window into the formation and evolution of massive stars and the compact remnants they leave behind \cite{LIGO_2016b}. This detection has been followed by 92 other events, significantly advancing our knowledge of the properties and behaviors of compact binaries \cite{LIGO_2016_O1,LVK_gwtc1_2019,  LVK_gwtc2_2021, LVK_gwtc21_2021, LVK_gwtc3_catalogue_2021}. Among them the community have identified 90 binary black holes (BBHs), 2 binary neutron stars (BNS) \cite{LIGO_GW170817,LIGO_GW190425}, and 2 neutron star-black hole (NSBHs) systems \cite{LIGO_NSBH_2021}, thereby offering new opportunities to study stellar astrophysics in ways that were previously inaccessible \cite{LVK_gwtc3_population_2021}.

Current gravitational wave observatories, LIGO\cite{LIGO_AdLIGO_2015}, Virgo\cite{Virgo_adVirgo_2015}, and KAGRA\cite{KAGRA_2018} (LVK) operate in the 10 Hz to several kHz frequency range, which is well-suited to observing the late stages of compact object binary coalescences. These observatories have already provided valuable data on black holes and neutron stars, including their masses, spins, and merger rates \cite{LVK_gwtc21_2021, LVK_gwtc3_population_2021}. The detections have shed light on the physics of massive stars, including their core-collapse processes, supernova mechanisms, and the formation of compact objects through both isolated and dynamical formation channels \cite{Mapelli_2021, Dominik_2012, Rodriguez_2016}.

Looking forward, next-generation terrestrial detectors such as the Einstein Telescope (ET) \cite{Coba} and Cosmic Explorer (CE) \cite{CE} are expected to significantly enhance our understanding of stellar astrophysics by extending the detection horizon to redshifts z =100 and lower masses \cite{ET, CE}. These instruments will be capable of detecting gravitational waves from the mergers of compact objects formed in the early universe, providing insights into the evolution of stellar populations at cosmic dawn \cite{Wang_2022, Santoliquido_2023}. Importantly, they will also provide valuable insights into the potential existence and formation of intermediate-mass black holes (IMBHs), whose origins, if they exist, remain uncertain but are thought to be possibly linked to the evolution of very massive stars or dense stellar environments such as star clusters \cite{Amaro-Seoane_2009, Jani_2019, Liu_2024, Mestichelli_2024, Vaccaro_2023}.

The motivation for studying gravitational waves from compact object mergers is deeply rooted in the desire to understand stellar evolution. Compact objects—black holes and neutron stars—are the final stages in the lives of massive stars. Their masses, spins, and the rates at which they merge are directly tied to stellar properties such as mass loss, metallicity, and binarity \cite{Belczynski_2016, Mapelli_2019,Giacobbo_2018}. For instance, the discovery of heavy stellar-mass black holes with masses exceeding 30 solar masses challenges traditional models of stellar evolution and supernova theory, suggesting the need to reconsider how massive stars evolve and end their lives \cite{Woosley_2016, Kroupa_2021, Sana_2012, Bavera_2022}. Similarly, neutron star mergers provide crucial information about the internal composition of neutron stars and the equation of state of ultra-dense matter, which remains one of the key unknowns in nuclear astrophysics \cite{LIGO_GW170817, Annala_2017}.

To further advance our understanding of these processes, we present \princess{} a novel predictive tool designed to simulate and forecast gravitational wave observations from user catalogue. This tool can predict both individual compact object mergers and the associated astrophysical gravitational wave background, which arises from the cumulative contributions of unresolved binary mergers throughout cosmic history \cite{Bavera_2021, Perigois_2020, Perigois_2022, Liu_2024}. 

In this paper, we detail the usage and functionnality of \princess{} as well as all the theroretical compotations hidden in the program. The rest of the paper is organized as follow: The first section details the structure of the code as well as a guide for a first basic use. The second section descibes all theoretical assumptions in the computations. The third section propose an example of predictions made with \princess{} based on three astrophysical models. The paper concludes with a last section gathering the main idea of the \princess{} structure, the main results from our predictions and discussion around future upgrades.

\section{Quick Start}
\label{sec:quickstart}

This section describes the structure of the code, the main steps of the computations, and contains instructions for launching a basic computation.

\subsection{Installation and initial checks}
\label{sec:install_and_checks}
\princess{} is a public code available on GitHub\footnote{\href{https://github.com/Cperigois/Princess}{https://github.com/Cperigois/Princess}} and GitLab\footnote{\href{https://gitlab.com/Cperigois/Princess}{https://gitlab.com/Cperigois/Princess}}. For first-time users, the directory can be easily cloned. We recommend that users install the following Python packages and check their versions before running the code:

\begin{itemize}
	\item Python 3.7 or higher
	\item PyCBC v1.18 (more recent versions are \textbf{NOT} compatible with \princess{})\cite{Pycbc-2005, Pycbc-2014, Pycbc-2017}
	\item Pandas \cite{Pandas}
	\item NumPy \cite{Numpy}
	\item SciPy \cite{SciPy_2020}
\end{itemize}

In the main directory of \texttt{Princess}, there is a script called \texttt{test.py}. We recommend running this test before starting any analysis. This script checks the versions of all the packages and performs a quick run of \princess{} on a mini-catalog to verify that all basic functionalities are working correctly. If the user has everything installed properly but encounters another issue, they are encouraged to contact the corresponding author. \princess{} is a program under development, and recent changes may affect the versions available on GitHub and GitLab.

\subsection{Structure of the repository}
\label{sec:structure}

The files downloaded from GitHub or GitLab have the following structure:

\dirtree{%
.1 Princess.
.2 AuxiliaryFiles.
.2 Run.
.3 \textbf{advanced\_params.py}.
.3 \textbf{getting\_started.py}.
.2 astrotools.
.2 gwtools.
.2 stochastic.
.2 README.md.
.2 \textbf{run.py}.
}

For basic computations, the user needs to be familiar with and check three files: 

\begin{itemize}
    \item \textbf{\texttt{run.py}} is the code that executes the full analysis. This code should \textbf{not} be modified. Section \ref{sec:run.py} details all the different steps of the program.
    
    \item \textbf{\texttt{Run/getting\_started.py}} is the first file to open and fill out for a computation. An example of how to complete this file can be found in Section \ref{sec:basic_settings}.
    
    \item \textbf{\texttt{Run/advanced\_params.py}} contains all the default parameters preset for an analysis. In this file, users can customize computations, such as changing frequency bounds or detector configurations. Section \ref{app:advparam} provides details about these parameters and their use in the program. 
\end{itemize}

The \texttt{astrotools}, \texttt{gwtools}, and \texttt{stochastic} packages contain all the classes, methods, and functions related to binary compact object astrophysics, gravitational wave computations, and stochastic background computations, respectively. The \texttt{AuxiliaryFiles} directory contains all data related to the detectors and previous detections. Section \ref{app:auxfiles} gathers all files and links to the original publications of the data.

\subsection{\texttt{run.py}}
\label{sec:run.py}
This section details the various steps of computation encapsulated in the functions defined in the file \texttt{run.py}.

\subsubsection{Initialization}
\label{sec:init}

  \begin{tcolorbox}[breakable, size=fbox, boxrule=1pt, pad at break*=1mm, colback=cellbackground, colframe=cellborder]
\prompt{In}{incolor}{1}{\boxspacing}
\begin{Verbatim}[commandchars=\\\{\}]
\PY{k+kn}{import} \PY{n+nn}{os}
\PY{k+kn}{import} \PY{n+nn}{json}
\PY{k+kn}{import} \PY{n+nn}{Run}\PY{n+nn}{.}\PY{n+nn}{getting\PYZus{}started} \PY{k}{as} \PY{n+nn}{GS} \PY{c+c1}{\PYZsh{} Initiate the file Params.json}
\PY{k+kn}{import} \PY{n+nn}{Run}\PY{n+nn}{.}\PY{n+nn}{advanced\PYZus{}params} \PY{k}{as} \PY{n+nn}{AP}
\PY{k+kn}{import} \PY{n+nn}{astrotools}\PY{n+nn}{.}\PY{n+nn}{astromodel} \PY{k}{as} \PY{n+nn}{AM}
\PY{k+kn}{import} \PY{n+nn}{astrotools}\PY{n+nn}{.}\PY{n+nn}{detection} \PY{k}{as} \PY{n+nn}{DET}
\PY{k+kn}{import} \PY{n+nn}{stochastic}\PY{n+nn}{.}\PY{n+nn}{background} \PY{k}{as} \PY{n+nn}{BKG}
\end{Verbatim}
\end{tcolorbox}


    \begin{tcolorbox}[breakable, size=fbox, boxrule=1pt, pad at break*=1mm, colback=cellbackground, colframe=cellborder]
\prompt{In}{incolor}{2}{\boxspacing}
\begin{Verbatim}[commandchars=\\\{\}]
\PY{n}{params} \PY{o}{=} \PY{n}{json}\PY{o}{.}\PY{n}{load}\PY{p}{(}\PY{n+nb}{open}\PY{p}{(}\PY{l+s+s1}{\PYZsq{}}\PY{l+s+s1}{Run/Params.json}\PY{l+s+s1}{\PYZsq{}}\PY{p}{,} \PY{l+s+s1}{\PYZsq{}}\PY{l+s+s1}{r}\PY{l+s+s1}{\PYZsq{}}\PY{p}{)}\PY{p}{)}
\end{Verbatim}
\end{tcolorbox}

In these two first steps, the script creates the file \texttt{Params.json}, which gathers all parameters set in \texttt{getting\_started.py} and \texttt{advanced\_params.py}. The \texttt{Params.json} file will subsequently be loaded in all further steps.


    \begin{tcolorbox}[breakable, size=fbox, boxrule=1pt, pad at break*=1mm, colback=cellbackground, colframe=cellborder]
\prompt{In}{incolor}{3}{\boxspacing}
\begin{Verbatim}[commandchars=\\\{\}]
\PY{k}{if} \PY{o+ow}{not} \PY{n}{os}\PY{o}{.}\PY{n}{path}\PY{o}{.}\PY{n}{exists}\PY{p}{(}\PY{l+s+s1}{\PYZsq{}}\PY{l+s+s1}{Run/}\PY{l+s+s1}{\PYZsq{}} \PY{o}{+} \PY{n}{params}\PY{p}{[}\PY{l+s+s1}{\PYZsq{}}\PY{l+s+s1}{name\PYZus{}of\PYZus{}project\PYZus{}folder}\PY{l+s+s1}{\PYZsq{}}\PY{p}{]}\PY{p}{)}\PY{p}{:}
    \PY{n}{os}\PY{o}{.}\PY{n}{mkdir}\PY{p}{(}\PY{l+s+s1}{\PYZsq{}}\PY{l+s+s1}{Run/}\PY{l+s+s1}{\PYZsq{}} \PY{o}{+} \PY{n}{params}\PY{p}{[}\PY{l+s+s1}{\PYZsq{}}\PY{l+s+s1}{name\PYZus{}of\PYZus{}project\PYZus{}folder}\PY{l+s+s1}{\PYZsq{}}\PY{p}{]}\PY{p}{)}
\PY{k}{if} \PY{o+ow}{not} \PY{n}{os}\PY{o}{.}\PY{n}{path}\PY{o}{.}\PY{n}{exists}\PY{p}{(}\PY{l+s+s1}{\PYZsq{}}\PY{l+s+s1}{Run/}\PY{l+s+s1}{\PYZsq{}} \PY{o}{+} \PY{n}{params}\PY{p}{[}\PY{l+s+s1}{\PYZsq{}}\PY{l+s+s1}{name\PYZus{}of\PYZus{}project\PYZus{}folder}\PY{l+s+s1}{\PYZsq{}}\PY{p}{]} \PY{o}{+} \PY{l+s+s2}{\PYZdq{}}\PY{l+s+s2}{/Astro\PYZus{}Models/}\PY{l+s+s2}{\PYZdq{}}\PY{p}{)}\PY{p}{:}
    \PY{n}{os}\PY{o}{.}\PY{n}{mkdir}\PY{p}{(}\PY{l+s+s1}{\PYZsq{}}\PY{l+s+s1}{Run/}\PY{l+s+s1}{\PYZsq{}} \PY{o}{+} \PY{n}{params}\PY{p}{[}\PY{l+s+s1}{\PYZsq{}}\PY{l+s+s1}{name\PYZus{}of\PYZus{}project\PYZus{}folder}\PY{l+s+s1}{\PYZsq{}}\PY{p}{]} \PY{o}{+} \PY{l+s+s2}{\PYZdq{}}\PY{l+s+s2}{/Astro\PYZus{}Models/}\PY{l+s+s2}{\PYZdq{}}\PY{p}{)}
\PY{k}{if} \PY{o+ow}{not} \PY{n}{os}\PY{o}{.}\PY{n}{path}\PY{o}{.}\PY{n}{exists}\PY{p}{(}\PY{l+s+s1}{\PYZsq{}}\PY{l+s+s1}{Run/}\PY{l+s+s1}{\PYZsq{}} \PY{o}{+} \PY{n}{params}\PY{p}{[}\PY{l+s+s1}{\PYZsq{}}\PY{l+s+s1}{name\PYZus{}of\PYZus{}project\PYZus{}folder}\PY{l+s+s1}{\PYZsq{}}\PY{p}{]} \PY{o}{+} \PY{l+s+s2}{\PYZdq{}}\PY{l+s+s2}{/Astro\PYZus{}Models/Catalogs/}\PY{l+s+s2}{\PYZdq{}}\PY{p}{)}\PY{p}{:}
    \PY{n}{os}\PY{o}{.}\PY{n}{mkdir}\PY{p}{(}\PY{l+s+s1}{\PYZsq{}}\PY{l+s+s1}{Run/}\PY{l+s+s1}{\PYZsq{}} \PY{o}{+} \PY{n}{params}\PY{p}{[}\PY{l+s+s1}{\PYZsq{}}\PY{l+s+s1}{name\PYZus{}of\PYZus{}project\PYZus{}folder}\PY{l+s+s1}{\PYZsq{}}\PY{p}{]} \PY{o}{+} \PY{l+s+s2}{\PYZdq{}}\PY{l+s+s2}{/Astro\PYZus{}Models/Catalogs}\PY{l+s+s2}{\PYZdq{}}\PY{p}{)}
\end{Verbatim}
\end{tcolorbox}

\textcolor{incolor}{\texttt{[3]}} prepares directories to store all outputs from the project. The details of the outputs are discussed in Section \ref{sec:output}.


    \begin{tcolorbox}[breakable, size=fbox, boxrule=1pt, pad at break*=1mm, colback=cellbackground, colframe=cellborder]
\prompt{In}{incolor}{4}{\boxspacing}
\begin{Verbatim}[commandchars=\\\{\}]
\PY{c+c1}{\PYZsh{} Read and reshape astrophysical models, saving instances of each for later use}
\PY{n}{AM}\PY{o}{.}\PY{n}{initialization}\PY{p}{(}\PY{p}{)}

\PY{c+c1}{\PYZsh{} Read and reshape detectors and networks, saving instances for each of them}
\PY{n}{DET}\PY{o}{.}\PY{n}{initialization}\PY{p}{(}\PY{p}{)}
\end{Verbatim}
\end{tcolorbox}


The functions \texttt{AM.initialization()} and \texttt{DET.initialization()} aim to create and save instances of classes for all astrophysical models, detectors, and networks. This allows the user to access temporary files (\texttt{.pickle}) that save time for subsequent computations. These \texttt{.pickle} files are removed at the end of the program using the function \texttt{AP.clean()}.

\subsubsection{Individual detections}
\label{sec:individual_detections}

    \begin{tcolorbox}[breakable, size=fbox, boxrule=1pt, pad at break*=1mm, colback=cellbackground, colframe=cellborder]
\prompt{In}{incolor}{5}{\boxspacing}
\begin{Verbatim}[commandchars=\\\{\}]
\PY{c+c1}{\PYZsh{} Compute SNR and conduct individual analyses for each detection}
\PY{n}{AM}\PY{o}{.}\PY{n}{process\PYZus{}astromodel}\PY{p}{(}\PY{p}{)}
\end{Verbatim}
\end{tcolorbox}

By calling the method \texttt{process\_astromodel()}, the program processes each existing \textit{Astromodel} instance through the following steps:
\begin{itemize}
    \item \texttt{make\_catalog()}: Reads and reshapes the catalog. This reshaping includes normalizing column names as specified in the file \texttt{advanced\_params.py}, and computing additional quantities such as redshifts, distances, or spins. The details of spin computation and its options are discussed in Section \ref{sec:GWwaveform}. The computations of distances and redshifts are performed using the astropy module \cite{astropy}.
    \item \texttt{compute\_SNR()}: Computes the individual SNR for each binary. The SNR computation is carried out through waveform evaluation, as detailed in the theoretical section \ref{sec:detectors_and_detectability}.
\end{itemize}
This step concludes with the creation of catalogs named after the corresponding astrophysical model. These catalogs contain all binaries from the initial input along with their individual SNR values. More details about these files can be found in Section \ref{sec:output}.


\subsubsection{Background computation}
\label{sec:background_computation}


    \begin{tcolorbox}[breakable, size=fbox, boxrule=1pt, pad at break*=1mm, colback=cellbackground, colframe=cellborder]
\prompt{In}{incolor}{6}{\boxspacing}
\begin{Verbatim}[commandchars=\\\{\}]
\PY{c+c1}{\PYZsh{} Compute backgrounds, residuals, and the corresponding analyses}
\PY{n}{BKG}\PY{o}{.}\PY{n}{process\PYZus{}background\PYZus{}computation}\PY{p}{(}\PY{p}{)}
\end{Verbatim}
\end{tcolorbox}


The background computation is currently performed by summing all individual sources from the input catalogs. The method \texttt{process\_background\_computation()} automatically computes the total and residual backgrounds\footnote{The residual background is the background expected after the perfect subtraction of all detected sources.}. It also calculates the SNR and the required observation time necessary for potential detection. Further details regarding spectral properties and background detection can be found in Section \ref{sec:GWbkg}.

\subsection*{}

    \begin{tcolorbox}[breakable, size=fbox, boxrule=1pt, pad at break*=1mm, colback=cellbackground, colframe=cellborder]
\prompt{In}{incolor}{7}{\boxspacing}
\begin{Verbatim}[commandchars=\\\{\}]
\PY{k}{if} \PY{n}{params}\PY{p}{[}\PY{l+s+s1}{\PYZsq{}}\PY{l+s+s1}{results}\PY{l+s+s1}{\PYZsq{}}\PY{p}{]}\PY{p}{[}\PY{l+s+s1}{\PYZsq{}}\PY{l+s+s1}{cleaning}\PY{l+s+s1}{\PYZsq{}}\PY{p}{]} \PY{o}{==} \PY{k+kc}{True} \PY{p}{:}
    \PY{n}{AP}\PY{o}{.}\PY{n}{clean}\PY{p}{(}\PY{p}{)}
\end{Verbatim}
\end{tcolorbox}

Finally, the run concludes with the removal of all temporary files that are not needed by the user, such as the \texttt{.pickle} files.

\textbf{NOTE: At this stage, a copy of the file \texttt{Params.json} located in the repository \texttt{/Run} is also removed. If, for any reason, the script does not complete and perform the cleaning, the user will need to manually remove the file \texttt{Run/Params.json}; otherwise, the parameter file will not be updated in subsequent runs.}

\subsection{Example: Basic settings}
\label{sec:basic_settings}
This section provides an example of predictions based on two catalogs of merging black holes and an Einstein Telescope (ET) arranged in a triangular configuration.

\begin{tcolorbox}[breakable, size=fbox, boxrule=1pt, pad at break*=1mm, colback=cellbackground, colframe=cellborder]
\prompt{In}{incolor}{1}{\boxspacing}
\begin{Verbatim}[commandchars=\\\{\}]
\PY{l+s+sd}{\PYZdq{}\PYZdq{}\PYZhy{}\PYZhy{}\PYZhy{}\PYZhy{}\PYZhy{}\PYZhy{}\PYZhy{}\PYZhy{}\PYZhy{}\PYZhy{}\PYZhy{}\PYZhy{}\PYZhy{}\PYZhy{}\PYZhy{}\PYZhy{}\PYZhy{}\PYZhy{}\PYZhy{}\PYZhy{}\PYZhy{}\PYZhy{}\PYZhy{}TO FILL\PYZhy{}\PYZhy{}\PYZhy{}\PYZhy{}\PYZhy{}\PYZhy{}\PYZhy{}\PYZhy{}\PYZhy{}\PYZhy{}\PYZhy{}\PYZhy{}\PYZhy{}\PYZhy{}\PYZhy{}\PYZhy{}\PYZhy{}\PYZhy{}\PYZhy{}\PYZhy{}\PYZhy{}\PYZhy{}\PYZdq{}\PYZdq{}\PYZdq{}}
\end{Verbatim}
\end{tcolorbox}

The user must carefully fill in the file \texttt{getting\_started.py} by following these steps:
\begin{itemize}[label = ]
    \item $\!$\prompt{In}{incolor}{1}{} \\ At this stage, the user chooses a name for the \textbf{project}/computation: \texttt{name\_of\_project\_folder}. This string will be used to create a repository in the folder \texttt{/Run/}, which will contain all outputs at the end of the computation. The output files are described in the next section \ref{sec:output}. The variable \texttt{n\_cpu\_max} is used to prevent overuse of the computer's capacity.
    
    \item $\!$\prompt{In}{incolor}{2}{} \\ The next step is to set and link the \textbf{astrophysical models} to be used. The user needs to fill in the dictionaries \texttt{astromodel\_1} and \texttt{astromodel\_2} with at least a name for each model (\texttt{model1} and \texttt{model2} in the example), a path to the catalog \texttt{original\_path}, and the observation duration of the catalog in years \texttt{duration}. Finally, all models need to be listed in the dictionary \texttt{astro\_model\_list}, as shown in the example. To add a third model, the user can create a dictionary called \texttt{astromodel\_3}, fill it with a \texttt{name}, an \texttt{original\_path}, and a \texttt{duration}, and then include the following entry in \texttt{astro\_model\_list}: \texttt{astromodel\_3['name']: astromodel\_3}. At this stage, it is also common to choose a spin model. By default, the program assumes that spins are included in the original files, but the user can choose an alternative option. The various spin options are detailed in section \ref{app:spin_options}.

    \item $\!$\prompt{In}{incolor}{3}{} \\ The next part to fill out concerns \textbf{detectors}. Similar to the astrophysical models, the user needs to set the detectors used in the computations using dictionaries, one for each detector. The mandatory dictionary entries for a detector include: a \texttt{name}, an \texttt{origin} indicating where to find the power spectral densities (PSD), a geometric \texttt{configuration}, a \texttt{psd\_file} which is the name of the file containing the PSD, and a \texttt{type} of detector, which can be either 2G or 3G. The \texttt{origin} variable must be chosen from \textbf{'Princess'} to use a preset PSD (a list of preset \texttt{psd\_file}s is available in Table~\ref{tab:PSD_princess}), \textbf{'Pycbc'} to use a sensitivity from \href{https://pycbc.org/pycbc/latest/html/pycbc.psd.html}{Pycbc}, where the \texttt{psd\_file} will be the reference name from Pycbc, or \textbf{'user'} for a personalized PSD, in which case the value of \texttt{psd\_file} should be the path to the user's PSD file. Finally, as with the models, all detectors must be included in the dictionary \texttt{detector\_list}. The network setup also follows the same structure in a dictionary, which includes a \texttt{name}, a \texttt{compo}sition (a dictionary of detectors comprising the network), a link to the power integrated curves (PIC) files (\texttt{pic\_file}), an efficiency value representing the proportion of time the detectors are in science mode together, and a signal-to-noise ratio (SNR) threshold \texttt{SNR\_thrs} defining the minimum SNR of an individual source to be considered resolved. The variable \texttt{frequency\_size} is an integer that defines the size of the frequency array for the computation. Finally, \texttt{rerun\_snr\_computation} and \texttt{rerun\_detectors} are boolean variables that allow the user to skip the computation of individual SNRs and the PSDs if they have already been computed.

    \item $\!$\prompt{In}{incolor}{4}{} \\ The next three boolean variables are used to enable or disable the final steps of computation, such as the cleaning process that removes all unnecessary temporary files generated during the computation and the plotting part, which displays basic plots of the backgrounds of each model and the network's sensitivity.

    \item $\!$\prompt{In}{incolor}{5}{} \\ This last part of the script should NOT be modified. It contains the instructions to build the \texttt{Params.json} file.
\end{itemize}
This script showcases only the most variable parameters from one project to another. The file \texttt{/Run/advanced\_params.py} contains all hidden default parameters, such as frequency bounds for different types of detectors, dictionaries of labels to read the input catalogs, etc. A specific appendix \ref{app:advparam} will be dedicated to enumerating all the variables found inside.


    \begin{tcolorbox}[breakable, size=fbox, boxrule=1pt, pad at break*=1mm,colback=cellbackground, colframe=cellborder]
\prompt{In}{incolor}{1}{\boxspacing}
\begin{Verbatim}[commandchars=\\\{\}]
\PY{l+s+sd}{\PYZdq{}\PYZdq{}\PYZdq{}             *** GENERIC PARAMETERS ***            \PYZdq{}\PYZdq{}\PYZdq{}}

\PY{n}{name\PYZus{}of\PYZus{}project\PYZus{}folder} \PY{o}{=} \PY{l+s+s1}{\PYZsq{}}\PY{l+s+s1}{Project1}\PY{l+s+s1}{\PYZsq{}}
\PY{n}{n\PYZus{}cpu\PYZus{}max} \PY{o}{=} \PY{l+m+mi}{4}  \PY{c+c1}{\PYZsh{} Number maximal of cpu used by the code}
\PY{n}{param\PYZus{}dictionary} \PY{o}{=} \PY{p}{\PYZob{}}\PY{l+s+s1}{\PYZsq{}}\PY{l+s+s1}{name\PYZus{}of\PYZus{}project\PYZus{}folder}\PY{l+s+s1}{\PYZsq{}}\PY{p}{:} \PY{n}{name\PYZus{}of\PYZus{}project\PYZus{}folder}\PY{p}{\PYZcb{}}
\PY{n}{AP}\PY{o}{.}\PY{n}{set}\PY{p}{(}\PY{n}{name\PYZus{}of\PYZus{}project\PYZus{}folder}\PY{p}{,} \PY{n}{param\PYZus{}dictionary}\PY{p}{,} \PY{n}{AP}\PY{o}{.}\PY{n}{advParams}\PY{p}{)}
\end{Verbatim}
\end{tcolorbox}


\begin{tcolorbox}[breakable, size=fbox, boxrule=1pt, pad at break*=1mm,colback=cellbackground, colframe=cellborder]
\prompt{In}{incolor}{2}{\boxspacing}
\begin{Verbatim}[commandchars=\\\{\}]
\PY{l+s+sd}{\PYZdq{}\PYZdq{}\PYZdq{}               *** ASTROMODELS ***                 \PYZdq{}\PYZdq{}\PYZdq{}}
\PY{l+s+sd}{\PYZdq{}\PYZdq{}\PYZdq{}}
\PY{n}{path} \PY{o}{=}  \PY{l+s+s1}{\PYZsq{}}\PY{l+s+s1}{/home/carole/Documents/PRINCESS\PYZus{}tutorial/}\PY{l+s+s1}{\PYZsq{}}
\PY{n}{astromodel\PYZus{}1} \PY{o}{=} \PY{p}{\PYZob{}}\PY{l+s+s1}{\PYZsq{}}\PY{l+s+s1}{name}\PY{l+s+s1}{\PYZsq{}}\PY{p}{:} \PY{l+s+s1}{\PYZsq{}}\PY{l+s+s1}{model1}\PY{l+s+s1}{\PYZsq{}}\PY{p}{,}
                \PY{l+s+s1}{\PYZsq{}}\PY{l+s+s1}{original\PYZus{}path}\PY{l+s+s1}{\PYZsq{}}\PY{p}{:} \PY{n}{path}\PY{o}{+}\PY{l+s+s1}{\PYZsq{}}\PY{l+s+s1}{Catalog\PYZus{}1.dat}\PY{l+s+s1}{\PYZsq{}}\PY{p}{,}
                \PY{l+s+s1}{\PYZsq{}}\PY{l+s+s1}{spin\PYZus{}model}\PY{l+s+s1}{\PYZsq{}} \PY{p}{:} \PY{l+s+s1}{\PYZsq{}}\PY{l+s+s1}{Zeros}\PY{l+s+s1}{\PYZsq{}}\PY{p}{,}
                \PY{l+s+s1}{\PYZsq{}}\PY{l+s+s1}{duration}\PY{l+s+s1}{\PYZsq{}}\PY{p}{:} \PY{l+m+mi}{1}\PY{p}{\PYZcb{}}

\PY{n}{astromodel\PYZus{}2} \PY{o}{=} \PY{p}{\PYZob{}}\PY{l+s+s1}{\PYZsq{}}\PY{l+s+s1}{name}\PY{l+s+s1}{\PYZsq{}}\PY{p}{:} \PY{l+s+s1}{\PYZsq{}}\PY{l+s+s1}{model2}\PY{l+s+s1}{\PYZsq{}}\PY{p}{,}
                \PY{l+s+s1}{\PYZsq{}}\PY{l+s+s1}{original\PYZus{}path}\PY{l+s+s1}{\PYZsq{}}\PY{p}{:} \PY{n}{path}\PY{o}{+}\PY{l+s+s1}{\PYZsq{}}\PY{l+s+s1}{Catalog\PYZus{}2.dat}\PY{l+s+s1}{\PYZsq{}}\PY{p}{,}
                \PY{l+s+s1}{\PYZsq{}}\PY{l+s+s1}{spin\PYZus{}model}\PY{l+s+s1}{\PYZsq{}} \PY{p}{:} \PY{l+s+s1}{\PYZsq{}}\PY{l+s+s1}{Zeros}\PY{l+s+s1}{\PYZsq{}}\PY{p}{,}
                \PY{l+s+s1}{\PYZsq{}}\PY{l+s+s1}{duration}\PY{l+s+s1}{\PYZsq{}}\PY{p}{:} \PY{l+m+mi}{1}\PY{p}{\PYZcb{}}

\PY{n}{astro\PYZus{}model\PYZus{}list} \PY{o}{=} \PY{p}{\PYZob{}}\PY{n}{astromodel\PYZus{}1}\PY{p}{[}\PY{l+s+s1}{\PYZsq{}}\PY{l+s+s1}{name}\PY{l+s+s1}{\PYZsq{}}\PY{p}{]}\PY{p}{:} \PY{n}{astromodel\PYZus{}1}\PY{p}{,}
                    \PY{n}{astromodel\PYZus{}2}\PY{p}{[}\PY{l+s+s1}{\PYZsq{}}\PY{l+s+s1}{name}\PY{l+s+s1}{\PYZsq{}}\PY{p}{]}\PY{p}{:} \PY{n}{astromodel\PYZus{}2}\PY{p}{\PYZcb{}}
\end{Verbatim}
\end{tcolorbox}

\begin{tcolorbox}[breakable, size=fbox, boxrule=1pt, pad at break*=1mm,colback=cellbackground, colframe=cellborder]
\prompt{In}{incolor}{3}{\boxspacing}
\begin{Verbatim}[commandchars=\\\{\}]
\PY{l+s+sd}{\PYZdq{}\PYZdq{}\PYZdq{}               *** Detectors and Network ***                 \PYZdq{}\PYZdq{}\PYZdq{}}
\PY{l+s+sd}{\PYZdq{}\PYZdq{}\PYZdq{}}
\PY{l+s+sd}{        Set the runs you want to use}
\PY{l+s+sd}{        List of available detectors : }
\PY{l+s+sd}{\PYZdq{}\PYZdq{}\PYZdq{}}

\PY{n}{rerun\PYZus{}snr\PYZus{}computation} \PY{o}{=} \PY{k+kc}{False}

\PY{n}{frequency\PYZus{}size} \PY{o}{=} \PY{l+m+mi}{2500} \PY{c+c1}{\PYZsh{} need to be an int}

\PY{c+c1}{\PYZsh{} Define detectors}
\PY{n}{detector\PYZus{}1} \PY{o}{=} \PY{p}{\PYZob{}}\PY{l+s+s1}{\PYZsq{}}\PY{l+s+s1}{name}\PY{l+s+s1}{\PYZsq{}} \PY{p}{:} \PY{l+s+s1}{\PYZsq{}}\PY{l+s+s1}{ET}\PY{l+s+s1}{\PYZsq{}}\PY{p}{,} 
              \PY{l+s+s1}{\PYZsq{}}\PY{l+s+s1}{origin}\PY{l+s+s1}{\PYZsq{}}\PY{p}{:} \PY{l+s+s1}{\PYZsq{}}\PY{l+s+s1}{Pycbc}\PY{l+s+s1}{\PYZsq{}}\PY{p}{,} 
              \PY{l+s+s1}{\PYZsq{}}\PY{l+s+s1}{configuration}\PY{l+s+s1}{\PYZsq{}} \PY{p}{:} \PY{l+s+s1}{\PYZsq{}}\PY{l+s+s1}{ET}\PY{l+s+s1}{\PYZsq{}}\PY{p}{,} 
              \PY{l+s+s1}{\PYZsq{}}\PY{l+s+s1}{psd\PYZus{}file}\PY{l+s+s1}{\PYZsq{}} \PY{p}{:} \PY{l+s+s1}{\PYZsq{}}\PY{l+s+s1}{EinsteinTelescopeP1600143}\PY{l+s+s1}{\PYZsq{}}\PY{p}{,} 
              \PY{l+s+s1}{\PYZsq{}}\PY{l+s+s1}{type}\PY{l+s+s1}{\PYZsq{}} \PY{p}{:} \PY{l+s+s1}{\PYZsq{}}\PY{l+s+s1}{3G}\PY{l+s+s1}{\PYZsq{}}\PY{p}{\PYZcb{}}

\PY{n}{detector\PYZus{}list} \PY{o}{=} \PY{p}{\PYZob{}}\PY{n}{detector\PYZus{}1}\PY{p}{[}\PY{l+s+s1}{\PYZsq{}}\PY{l+s+s1}{name}\PY{l+s+s1}{\PYZsq{}}\PY{p}{]}\PY{p}{:} \PY{n}{detector\PYZus{}1}\PY{p}{\PYZcb{}}

\PY{l+s+s2}{\PYZdq{}}\PY{l+s+s2}{               ***                 }\PY{l+s+s2}{\PYZdq{}}
\PY{n}{network\PYZus{}1} \PY{o}{=} \PY{p}{\PYZob{}}\PY{l+s+s1}{\PYZsq{}}\PY{l+s+s1}{name}\PY{l+s+s1}{\PYZsq{}} \PY{p}{:} \PY{l+s+s1}{\PYZsq{}}\PY{l+s+s1}{ET\PYZus{}net}\PY{l+s+s1}{\PYZsq{}}\PY{p}{,}
             \PY{l+s+s1}{\PYZsq{}}\PY{l+s+s1}{compo}\PY{l+s+s1}{\PYZsq{}} \PY{p}{:} \PY{p}{\PYZob{}}\PY{n}{detector\PYZus{}1}\PY{p}{[}\PY{l+s+s1}{\PYZsq{}}\PY{l+s+s1}{name}\PY{l+s+s1}{\PYZsq{}}\PY{p}{]} \PY{p}{:} \PY{n}{detector\PYZus{}1}\PY{p}{\PYZcb{}}\PY{p}{,}
             \PY{l+s+s1}{\PYZsq{}}\PY{l+s+s1}{pic\PYZus{}file}\PY{l+s+s1}{\PYZsq{}} \PY{p}{:} \PY{l+s+s1}{\PYZsq{}}\PY{l+s+s1}{AuxiliaryFiles/PICs/ET.txt}\PY{l+s+s1}{\PYZsq{}}\PY{p}{,}
             \PY{l+s+s1}{\PYZsq{}}\PY{l+s+s1}{efficiency}\PY{l+s+s1}{\PYZsq{}} \PY{p}{:} \PY{l+m+mf}{0.5}\PY{p}{,}
             \PY{l+s+s1}{\PYZsq{}}\PY{l+s+s1}{SNR\PYZus{}thrs}\PY{l+s+s1}{\PYZsq{}} \PY{p}{:} \PY{l+m+mi}{12}
             \PY{p}{\PYZcb{}}

\PY{n}{network\PYZus{}list} \PY{o}{=} \PY{p}{\PYZob{}}\PY{n}{network\PYZus{}1}\PY{p}{[}\PY{l+s+s1}{\PYZsq{}}\PY{l+s+s1}{name}\PY{l+s+s1}{\PYZsq{}}\PY{p}{]}\PY{p}{:} \PY{n}{network\PYZus{}1}\PY{p}{\PYZcb{}}

\PY{n}{rerun\PYZus{}detectors} \PY{o}{=} \PY{k+kc}{False}
\end{Verbatim}
\end{tcolorbox}


    \begin{tcolorbox}[breakable, size=fbox, boxrule=1pt, pad at break*=1mm,colback=cellbackground, colframe=cellborder]
\prompt{In}{incolor}{4}{\boxspacing}
\begin{Verbatim}[commandchars=\\\{\}]
\PY{n}{rerun\PYZus{}background} \PY{o}{=} \PY{k+kc}{False}

\PY{n}{run\PYZus{}data\PYZus{}cleaning} \PY{o}{=} \PY{k+kc}{True}
\PY{n}{run\PYZus{}plots} \PY{o}{=} \PY{k+kc}{False}
\end{Verbatim}
\end{tcolorbox}


    \begin{tcolorbox}[breakable, size=fbox, boxrule=1pt, pad at break*=1mm,colback=cellbackground, colframe=cellborder]
\prompt{In}{incolor}{5}{\boxspacing}
\begin{Verbatim}[commandchars=\\\{\}]
\PY{l+s+sd}{\PYZdq{}\PYZdq{}\PYZdq{}        *** Main Code, should not change ***       \PYZdq{}\PYZdq{}\PYZdq{}}

\PY{l+s+sd}{\PYZdq{}\PYZdq{}\PYZdq{}  1\PYZhy{} Set the directory for all intermediate and definitive results  \PYZdq{}\PYZdq{}\PYZdq{}}

\PY{k}{if} \PY{o+ow}{not} \PY{n}{os}\PY{o}{.}\PY{n}{path}\PY{o}{.}\PY{n}{exists}\PY{p}{(}\PY{l+s+s1}{\PYZsq{}}\PY{l+s+s1}{Run/}\PY{l+s+s1}{\PYZsq{}} \PY{o}{+} \PY{n}{name\PYZus{}of\PYZus{}project\PYZus{}folder}\PY{p}{)}\PY{p}{:}
    \PY{n}{os}\PY{o}{.}\PY{n}{mkdir}\PY{p}{(}\PY{l+s+s1}{\PYZsq{}}\PY{l+s+s1}{Run/}\PY{l+s+s1}{\PYZsq{}} \PY{o}{+} \PY{n}{name\PYZus{}of\PYZus{}project\PYZus{}folder}\PY{p}{)}

\PY{l+s+sd}{\PYZdq{}\PYZdq{}\PYZdq{}  2\PYZhy{} Gather and save the parameter used in the study  \PYZdq{}\PYZdq{}\PYZdq{}}

\PY{n}{param\PYZus{}dictionary} \PY{o}{=} \PY{p}{\PYZob{}}\PY{l+s+s1}{\PYZsq{}}\PY{l+s+s1}{name\PYZus{}of\PYZus{}project\PYZus{}folder}\PY{l+s+s1}{\PYZsq{}}\PY{p}{:} \PY{n}{name\PYZus{}of\PYZus{}project\PYZus{}folder}\PY{p}{,}
                    \PY{l+s+s1}{\PYZsq{}}\PY{l+s+s1}{astro\PYZus{}model\PYZus{}list}\PY{l+s+s1}{\PYZsq{}}\PY{p}{:} \PY{n}{astro\PYZus{}model\PYZus{}list}\PY{p}{,}
                    \PY{l+s+s1}{\PYZsq{}}\PY{l+s+s1}{detector\PYZus{}list}\PY{l+s+s1}{\PYZsq{}}\PY{p}{:} \PY{n}{detector\PYZus{}list}\PY{p}{,}
                    \PY{l+s+s1}{\PYZsq{}}\PY{l+s+s1}{network\PYZus{}list}\PY{l+s+s1}{\PYZsq{}} \PY{p}{:} \PY{n}{network\PYZus{}list}\PY{p}{,}
                    \PY{l+s+s1}{\PYZsq{}}\PY{l+s+s1}{frequency\PYZus{}size}\PY{l+s+s1}{\PYZsq{}} \PY{p}{:} \PY{n}{frequency\PYZus{}size}\PY{p}{,}
                    \PY{l+s+s1}{\PYZsq{}}\PY{l+s+s1}{n\PYZus{}cpu\PYZus{}max}\PY{l+s+s1}{\PYZsq{}}\PY{p}{:} \PY{n}{n\PYZus{}cpu\PYZus{}max}\PY{p}{,}
                    \PY{l+s+s1}{\PYZsq{}}\PY{l+s+s1}{overwrite}\PY{l+s+s1}{\PYZsq{}}\PY{p}{:} \PY{p}{\PYZob{}}\PY{l+s+s1}{\PYZsq{}}\PY{l+s+s1}{astromodel}\PY{l+s+s1}{\PYZsq{}}\PY{p}{:} \PY{n}{rerun\PYZus{}snr\PYZus{}computation}\PY{p}{,}
                                  \PY{l+s+s1}{\PYZsq{}}\PY{l+s+s1}{detectors}\PY{l+s+s1}{\PYZsq{}}\PY{p}{:} \PY{n}{rerun\PYZus{}detectors}\PY{p}{\PYZcb{}}\PY{p}{,}
                    \PY{l+s+s1}{\PYZsq{}}\PY{l+s+s1}{results}\PY{l+s+s1}{\PYZsq{}}\PY{p}{:} \PY{p}{\PYZob{}}\PY{l+s+s1}{\PYZsq{}}\PY{l+s+s1}{cleaning}\PY{l+s+s1}{\PYZsq{}}\PY{p}{:} \PY{n}{run\PYZus{}data\PYZus{}cleaning}\PY{p}{,}
                                \PY{l+s+s1}{\PYZsq{}}\PY{l+s+s1}{plots}\PY{l+s+s1}{\PYZsq{}}\PY{p}{:} \PY{n}{run\PYZus{}plots}\PY{p}{\PYZcb{}}
                    \PY{p}{\PYZcb{}}
\PY{n}{AP}\PY{o}{.}\PY{n}{set}\PY{p}{(}\PY{n}{name\PYZus{}of\PYZus{}project\PYZus{}folder}\PY{p}{,} \PY{n}{param\PYZus{}dictionary}\PY{p}{,} \PY{n}{AP}\PY{o}{.}\PY{n}{advParams}\PY{p}{)}
\end{Verbatim}
\end{tcolorbox}

\subsection{Example: Outputs}
\label{sec:output}
This section presents the outputs generated by the program. In our example, the project name is \texttt{Project1}, so the program created a repository named \texttt{Project1} where all outputs are stored. Once the computation is complete, this folder can be moved to a different location without affecting the program's functionality.

The following directory tree highlights the output repository and its contents in bold:

\dirtree{%
.1 Princess.
.2 AuxiliaryFiles.
.2 Run.
.3 \textbf{Project1}.
.4 \textbf{Astro\_Models}.
.5 \textbf{Catalogs}.
.6 \textbf{model1}.
.6 \textbf{model2}.
.4 \textbf{Results}.
.5 \textbf{Analysis}.
.6 \textbf{model1}.
.6 \textbf{model2}.
.5 \textbf{Omega}.
.6 \textbf{model1}.
.6 \textbf{model2}.
.4 \textbf{Params.json}.
.3 advanced\_params.py.
.3 getting\_started.py.
.2 astrotools.
.2 gwtools.
.2 stochastic.
.2 README.md.
.2 run.py.
}

The directory \texttt{\textbf{Project1/Astro\_Models/Catalogs}} contains all catalogs reshaped during the computation. The header of these catalogs includes the following columns, separated by tabs and presented in this order:

\begin{itemize}
    \item \texttt{\textbf{z}}: The redshift of the source.
    \item \texttt{\textbf{Mc}}: The chirp mass of the binary, source frame, in \msun.
    \item \texttt{\textbf{q}}: The mass ratio of the binary, $q = m_2/m_1$, assuming $m_1 > m_2$.
    \item \texttt{\textbf{m1}}: The mass of the more massive object in the binary, source frame, in \msun.
    \item \texttt{\textbf{m2}}: The mass of the second object in the binary, source frame, in \msun.
    \item \texttt{\textbf{Dl}}: The luminosity distance of the source, in Mpc.
    \item \texttt{\textbf{chi1}}: The spin component of the first object.
    \item \texttt{\textbf{chi2}}: The spin component of the second object.
    \item \texttt{\textbf{costheta1}}: The spin orientation of the first object.
    \item\texttt{\textbf{costheta2}}: The spin orientation of the second object.
    \item \texttt{\textbf{s1}}: The spin projection of the first object along the binary's angular momentum.
    \item \texttt{\textbf{s2}}: The spin projection of the second object along the binary's angular momentum.
    \item \texttt{\textbf{chip}}: $\chi_{\rm p}$, the precessing spin of the binary (see Sec.~\ref{sec:GWwaveform} for details).
    \item \texttt{\textbf{chieff}}: $\chi_{\rm eff}$, the effective spin of the binary (see Sec.~\ref{sec:GWwaveform} for details).
    \item \texttt{\textbf{ET}}: The individual signal-to-noise ratio (SNR) in the detector named \texttt{ET} (see Sec.~\ref{sec:detectors_and_detectability} for details).
    \item \texttt{\textbf{ET\_net}}: The individual SNR in the network named \texttt{ET\_net} (see Sec.~\ref{sec:detectors_and_detectability} for details).
\end{itemize}

The \texttt{Results} directory contains two subfolders: \texttt{Omega} and \texttt{Analysis}. The \texttt{Omega} folder stores the results of the background computation for each model. Each file contains the following columns, in order: \textbf{f}, the frequency in Hz, \textbf{Total}, representing the total background spectrum, and additional columns corresponding to the names of the networks, showing the residual background spectra.

The \texttt{Results/Analysis} folder includes the reference values of the computation, such as the number of sources in both the total and residual backgrounds, the values of Omega at 10 Hz and 25 Hz, and the SNR for both the total and residual backgrounds.

Finally, the output folder contains a copy of the \texttt{Params.json} file used for the computation.

\subsection{Common issues}
\label{sec:issues}

\textbf{Params.json not updating:}  
If you notice that certain changes made in the \texttt{getting\_started} file are not reflected in the \texttt{Params.json} file, this could indicate an issue with the update process.

\textbf{Changes with respect to initial catalogs:}
\princess{} does not compute or change any value from the initial catalog. The programm may compute additionnal variable such as the chirp mass. If you find any change with respect to your initial values please it may come from the shape of your initial file with is not correctly read by the programm. \princess{} make use of the function \texttt{read\_csv} from Pandas \cite{Pandas}. 

\textbf{Computational support:}  
If you have any issue or specific computation not working, please contact the author : caroleperigois@outlook.com

\section{Theoretical computations}
\label{sec:theory}

In this section, we detail the theoretical assessement behind the \princess{} code.

\subsection{Gravitational wave waveforms}
\label{sec:GWwaveform}
The computation of individual signal-to-noise ratios (SNRs) and the background spectrum requires preliminary calculations of the gravitational waveforms in the frequency domain. In \princess{}, there are two methods to compute these waveforms: analytical approximations from \cite{Ajith_2007, Ajith_2009} and the PyCBC software.

The GW waveform of a binary system $k$ is given by the sum of the squared Fourier domain GW amplitudes of the two polarizations $+$ and $\times$:

\begin{equation}
    \tilde{h}^2_{k}(f) = \tilde{h}^2_{+,k}(f) + \tilde{h}^2_{\times,k}(f).
\end{equation}

These contributions can be expressed as the product of the amplitude along the $z$-axis\footnote{Defined as orthogonal to the $x$ and $y$ axes carried by the two detector arms.} $h_{z,k}$, a factor depending on the binary inclination $\iota_k$, and $\Gamma$, which encodes the stages of coalescence:

\begin{align}
&\tilde{h}_{+,k}(f) = h_{z,k} \frac{1 + \cos^2(\iota_k)}{2} \Gamma(f),\\  
&\tilde{h}_{\times,k}(f) = h_{z,k} \cos(\iota_k) \Gamma(f),
\end{align}

where

\begin{equation}
    h_{z,k} = \sqrt{\frac{5}{24}} \frac{[G \mathcal{M}^{(z)}_k]^{5/6}}{\pi^{2/3} c^{3/2} d_L(z_k)},
    \label{h_z}
\end{equation}

with $\mathcal{M}^{(z)} = \frac{(m_1 m_2)^{3/5}}{(m_1 + m_2)^{1/5}(1 + z)}$ being the corrected chirp mass, where $z$ is the redshift, $d_L$ is the luminosity distance, and $\iota$ is the inclination angle of the binary. The function $\Gamma(f)$ varies with the coalescence stages and consists of three phases: the inspiral phase at frequencies $f < f_{\mathrm{merg}}$, the merger phase between $f_{\mathrm{merg}}$ and $f_{\mathrm{ring}}$, and the ringdown phase between $f_{\mathrm{ring}}$ and $f_{\mathrm{cut}}$. For a circular orbit, 

\begin{equation}
\Gamma(f) = \left \{
\begin{array}{l l}
 (1 + \sum_{i=2}^{3} \alpha_i \nu^i) f^{-7/6} & \text{if } f < f_{\mathrm{merg}}  \\
 w_m (1 + \sum_{i=1}^{2} \epsilon_i \nu^i)^2 f^{-2/3} & \text{if } f_{\mathrm{merg}} \leq f < f_{\mathrm{ring}}  \\
 w_r \mathcal{L}^2(f, f_{\mathrm{ring}}, \sigma) & \text{if } f_{\mathrm{ring}} \leq f < f_{\mathrm{cut}}
 \end{array}
\right.
\text{,}
\end{equation}

with 

\begin{equation}
\begin{array}{l}
\nu \equiv (\pi Mf)^{1/3},\\
\epsilon_1 = 1.4547 \chi_{\mathrm{eff}} - 1.8897,\\
\epsilon_2 = -1.8153 \chi_{\mathrm{eff}} + 1.6557,\\
\alpha_2 = -\frac{323}{224} + \frac{451 \eta}{168}, \\
\alpha_3 = \left( \frac{27}{8} - \frac{11 \eta}{6} \right) \chi_{\mathrm{eff}},
\end{array}
\end{equation}

where $M = m_1 + m_2$ is the total mass of the binary and 

\begin{equation}
\chi_{\mathrm{eff}} = \frac{(\vec \chi_1 + q \vec \chi_2)}{1 + q} \cdot \frac{\vec L}{L}, 
\end{equation}

with $\vec \chi_{1,2} = \frac{\vec s_{1,2} c}{G m_{1,2}^2}$ being the dimensionless spin parameters of the two black holes.

The phenomenological expressions for the limit frequencies $\mu_k = [f_{\mathrm{merg}}, f_{\mathrm{ring}}, \sigma, f_{\mathrm{cut}}]$ are provided in \cite{Ajith_2009}:

\begin{equation}
\mu_k = \frac{1}{\pi M} \mu_k^0 + \sum_{i=1}^{3} \sum_{j=0}^{N} y_k^{(ij)} \eta^i \chi_{\mathrm{eff}}^j \,,
\label{eq:ampParams}
\end{equation}

\begin{table}[]
\hspace{-1.2cm}
\begin{tabular}{c|ccccccc}
\hline
               & Test-mass limit $(\mu^0_k)$                        & $y^{(10)}$ & $y^{(11)}$ & $y^{(12)}$ & $y^{(20)}$ & $y^{(21)}$ & $y^{(30)}$ \\ \hline
f$_{\mathrm{merg}}$ & $1 - 4.455(1 - \chi_{\mathrm{eff}})^{0.217} + 3.521(1 - \chi_{\mathrm{eff}})^{0.26}$ & 0.6437     & 0.827      & -0.2706    & -0.05822   & -3.935     & -7.092     \\
f$_{\mathrm{ring}}$ & $\frac{[1 - 0.63(1 - \chi_{\mathrm{eff}})^{0.3}]}{2}$                         & 0.1469     & -0.1228    & -0.02609   & -0.0249    & 0.1701     & 2.325      \\
$\sigma$       & $\frac{[1 - 0.63(1 - \chi_{\mathrm{eff}})^{0.3}] (1 - \chi_{\mathrm{eff}})^{0.45}}{4}$       & -0.4098    & -0.03523   & 0.1008     & 1.829      & -0.02017   & -2.87      \\
f$_{\mathrm{cut}}$  & $0.3236 + 0.04894 \chi_{\mathrm{eff}} + 0.01346 \chi_{\mathrm{eff}}^2$       & -0.1331    & -0.08172   & 0.1451     & -0.2714    & 0.1279     & 4.922      \\ \hline
\end{tabular}
\caption{Extraction from Table I of Ajith et al. 2011}
\end{table}

\subsection{Detectors and detectability}
\label{sec:detectors_and_detectability}

In this section, we discuss the computation of signal-to-noise ratios (SNRs) for a GW event arriving at a detector. Currently, \princess{} can be used for ground-based detectors only, some of which are preset and discussed in Appendix \ref{sec:detection}.

\paragraph{Optimal SNR: }The confidence in the detection of a GW event is quantified by three main variables: its astrophysical probability $p_{\rm astro}$, false alarm rate (FAR), and SNR $\rho$. \princess{} provides tools to calculate the SNR of individual events, denoted as $\rho^k$\footnote{To avoid confusion, the SNR for a single source $k$ will be written as $\rho^k$, while SNR will refer to the detectability of the background.}. For a single detector $d$, the optimal SNR is calculated under the assumption that the source is perfectly oriented and located in the sky. Thus, for an event $k$, $\rho$ can be expressed as:

\begin{equation}
    \rho_{opt, k}^2 = 4 \int_{f_{min}}^{f_{max}} \frac{\tilde{h}_k^2(f)}{S_{n,d}(f)} \mathrm{d}f,
\end{equation}

where $S_{n,d}$ is the noise spectral density of detector $d$, and $f_{min}$ and $f_{max}$ define the frequency range of interest for detector $d$. 

\paragraph{Realistic SNR: }The calculation of a realistic SNR for detector $d$ and event $k$ simplifies to the product of the optimal SNR $\rho_{opt,d}$ and a factor $f_d$ that depends on the antenna pattern of the detectors $F_+(\theta, \phi)$ and $F_\times(\theta, \phi)$, where $\theta$ and $\phi$ refer to the declination and right ascension angles, respectively, as well as the inclination $\iota$ of the source.

Taking these variables into account, the realistic SNR is given by

\begin{equation}
    \rho_{k}^2 = 4 \int_{f_{min}}^{f_{max}} \frac{\tilde{h}_{z,k}^2 \left((1 + \cos^2 \iota_k) F_+(\theta, \phi)\right)^2 + \tilde{h}_{z,k}^2 \left(\cos \iota_k F_\times(\theta, \phi)\right)^2}{S_{n,d}(f)} \mathrm{d}f,
\end{equation}

which can be simplified to:

\begin{equation}
    \rho_{k,d}(\iota, \theta, \phi) = f_d \, \rho_{opt,k,d}.
\end{equation}

The calculation of $f_d$, as previously presented in \cite{Gerosa_2017, Finn_1992}, is based on the expressions for the antenna pattern:

\begin{align}
    & F_+(\theta, \phi) = \sin \xi \frac{(1 + \cos^2 \theta) \cos 2\phi}{2}, \\
    & F_\times(\theta, \phi) = \sin \xi \cos \theta \sin 2\phi,
\end{align}

where $\theta$ and $\phi$ are the declination and right ascension angles, and $\xi$ is the angle between the detector arms. Therefore, $f_d$ can be expressed as:

\begin{equation}
    f_d = \sin^2 \xi \left| \frac{(1 + \cos^2 \iota_k)(1 + \cos^2 \theta) \cos 2\phi}{4} \right|^2 + \sin^2 \xi \left| \cos \iota_k \cos \theta \sin 2\phi \right|^2.
\end{equation}

To avoid computational expense, \princess{} pre-computes $f_d$ for various detectors (see Appendix \ref{sec:detection} for more details) and compiles them in the file \texttt{AuxiliaryFiles/factor\_table.dat}.

\paragraph{Network SNR: }The signal-to-noise ratio $\rho_{Net}$ in a network comprising $N$ detectors $d$ can be estimated by combining the SNRs from each detector:

\begin{equation}
    \rho_{Net}^2 = \sum_{d=1}^{N} \rho_d^2.
\end{equation}

\subsection{GW background spectral properties}
\label{sec:GWbkg}

The background is characterized by the dimensionless quantity \cite{Allen_1997, Phinney_2001, Regimbau_2014, Perigois_2020}: 

\begin{equation}
    \Omega_{\rm gw}(f) = \frac{1}{c\rho_c} f \mathcal{F}(f),
    \label{omg}
\end{equation}
where $\rho_c = \frac{3H_0^2 c^2}{8\pi G}$ is the critical energy density of the Universe, and $\mathcal{F}(f)$ is the gravitational wave (GW) flux arriving at a detector with the detector-frame frequency $f$:
\begin{equation}
    \mathcal{F}(f) = T^{-1} \sum_{k=1}^{N} \frac{1}{4\pi r^2} \frac{dE_{\rm gw}^k}{df}(f).
    \label{flux}
\end{equation}
In the above equation, the sum is taken over the $N$ individual merger contributions to the flux during an observation time $T$. By rewriting the energy density, we can express the spectrum $\Omega_{\rm gw}(f)$ in terms of frequency-domain waveforms: 

\begin{equation}
    \Omega_{\rm gw}(f) = \frac{\pi c^2}{2 G \rho_c} f T^{-1} \sum_{k=1}^{N} f^2 \tilde{h}_k^2(f).
\end{equation}

\princess{} computes two types of GW backgrounds: the total background, assuming the catalog represents all sources in the Universe, and the residual background, which is the background calculated after the complete and perfect subtraction of events detected by the user-defined networks.

\paragraph{Background detection}

Finally, we quantify the detectability of the background by computing its signal-to-noise ratio (SNR). The SNR for a background is calculated by cross-correlating signals from two different detectors, $i$ and $j$. Assuming the noises of the detectors are independent and uncorrelated, the common part of the signals refers to the GW background:

\begin{equation}
\text{SNR}_{ij} = \frac{3 H_0^2}{10 \pi^2} \sqrt{2T} \left[
\int_0^\infty df \frac{\gamma_{ij}^2(f) \Omega_{\rm gw}^2(f)}{f^6 P_i(f) P_j(f)} \right]^{1/2},
\label{eq:snrCC}
\end{equation}

where $\gamma_{ij}(f)$ is the normalized isotropic overlap reduction function (ORF), which accounts for the reduction in sensitivity due to the separation and relative orientation of the two detectors \cite{Flanagan_1993,Christensen_1992}. Here, $P_i(f)$ and $P_j(f)$ are the one-sided power spectral noise densities for detectors $i$ and $j$, and $T$ is the effective observation time. 

By combining different detector pairs within an $n$-detector network \cite{Flanagan_1993,Christensen_1992}, we obtain: 
\begin{equation}
    \text{SNR} = \left[ \sum_{i=1}^n \sum_{j>i} \text{SNR}_{ij}^2 \right]^{1/2}
\end{equation}
For example, in the HLV case, we can write
\begin{equation}
    \text{SNR}_{HLV} = [\text{SNR}_{HL}^2 + \text{SNR}_{HV}^2 + \text{SNR}_{LV}^2]^{1/2}.
\end{equation}

\paragraph{Total versus residual backgrounds}

The residual background is defined as the background spectrum composed of sources that are not individually resolved. In \princess{}, the user can set a threshold $\rho_{\rm thrs}$ for the safe detection of individual events. To quantify the reduction of the background, we define the ratio of the remaining energy density, $r_{\Omega}$, and the proportion of mergers remaining after the subtraction, $r_{\rm count}$:
 
\begin{equation}
r_{\Omega} = \frac{\Omega_{\rm GW,res}(f_{\rm ref})}{\Omega_{\rm GW,tot}(f_{\rm ref})}, \qquad r_{\rm count} = \frac{N_{\rm res}}{N_{\rm tot}}.
\label{eq:ratio}
\end{equation}

\section{Observation predictions: an example}
\label{sec:example}

In this section, we present predictions for the detection of individual binary black hole (BBH) mergers in the Hanford-Livingston-Virgo (HLV), Einstein Telescope (ET), and two Cosmic Explorer (CE) networks. We utilized three catalogs that incorporate a mixture of binary formation and evolution channels, corresponding to different metallicity distribution deviations: $\sigma_\mathrm{Z} = 0.2$, $\sigma_\mathrm{Z} = 0.3$, and $\sigma_\mathrm{Z} = 0.4$, associated with models B02, B03, and B04 from \cite{Mapelli_2021}.

\subsection{Observations in 2G detectors: Predictions and comparison with previous runs}
\label{predic_2G}

The power spectral densities (PSD) used for each run are documented in Table \ref{tab:PSD_princess} (for O4, we used 'Virgo\_90Mpc\_O4'). For each run, we renormalized the predictions to the number of events per year, taking into account the duration of each run and the efficiency\footnote{Proportion of available science data from the run. Data obtained from \href{https://www.gw-openscience.org/eventapi/}{Gravitational-wave Open Science Center}.} listed in Table \ref{tab:runs}.

Figure \ref{fig:detections} illustrates the rate of individual detections for the three BBH catalogs with $\sigma_\mathrm{Z} = 0.2$, $\sigma_\mathrm{Z} = 0.3$, and $\sigma_\mathrm{Z} = 0.4$. The black stars represent the actual number of detected BBH mergers with a false alarm rate (FAR) $< 2$, an astrophysical probability $p_\mathrm{astro} > 0.85$, and a network signal-to-noise ratio $\rho_\mathrm{Net} > 8$ \cite{GWOSC_2021, GWOSC_2023}. These values can be retrieved in Table \ref{tab:detections}, which also provides the merger rates for each model. The error bars shown in the plot are computed from multiple realizations over inclination and sky position.

We observe an overestimate of the number of detected events, attributable to two factors. First, the simplifications made in \princess{} do not perfectly reflect the operations of the detectors. Second, the overestimation arises from the properties of the population: specifically, the merger rate and mass distribution. The latest estimation of the local merger rate for BBHs from the LVK collaboration \cite{LVK_gwtc3_population_2021} is between 18 and 39 Gpc$^{-3}$yr$^{-1}$; these limits correspond to the models $\sigma_\mathrm{Z} = 0.2$ and $\sigma_\mathrm{Z} = 0.3$ (see Table \ref{tab:detections}). The persisting overestimation comes from the masses. Figure \ref{fig:Mc_distrib} shows the primary mass distribution for the three models and the last LVK results. The models are predicting an important number of mergers with a primary mass $m_1>15$\msun. this feature has an important impact on the snr of the sources which depends mostly on the masses and the distance. This overestimate of the number of bniary with large primary mass, is a well know issue for every model build before the O3a realese \cite{Broekgaarden_2021} and \textit{Sgalletta et al. in prep}.



Given these predictions, it seems reasonable to anticipate the detection of several hundred BBHs in the upcoming O4 run.

\begin{table}[]
\centering
\makebox[0pt][c]{\parbox{1\textwidth}{%
\begin{minipage}{0.5\textwidth}
\centering
    \includegraphics[width=8cm]{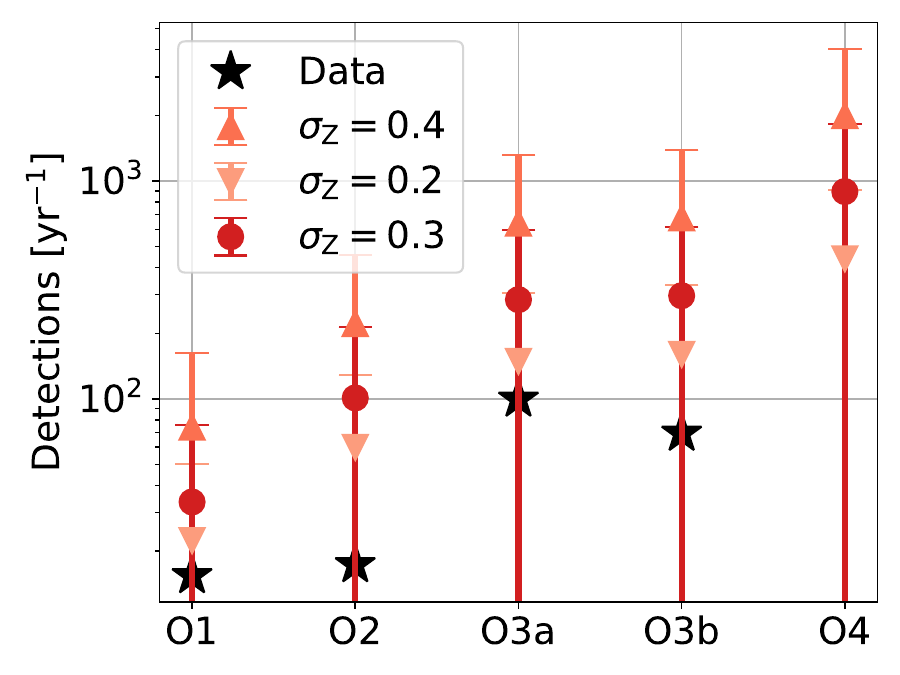}
    \captionof{figure}{Predicted BBH detections by \princess{} for three different models based on metallicity standard deviation. Error bars represent statistical fluctuations in inclination and sky position.}
    \label{fig:detections}
\end{minipage}
\hfill
\begin{minipage}{0.50\textwidth}
\centering
\begin{tabular}{c|c|c}
\hline
Run & Duration [days] & Efficiency \\ \hline
O1  & 129             & 0.55       \\
O2  & 268             & 0.55       \\
O3a & 183             & 0.70       \\
O3b & 147             & 0.75       \\
O4  & 365             & 0.75       \\ \hline
\end{tabular}
\caption{Summary of the runs used for the calculations.}
\label{tab:runs}
\end{minipage}}}
\end{table}

\begin{table}[]
\centering
\begin{tabular}{l|c|c|c|c|c|c}
\hline
                                 & \textbf{\begin{tabular}[c]{@{}c@{}}$\mathcal{R}_0$\\ Gpc$^{-3}$.yr$^{-1}$\end{tabular}} & \textbf{O1}   & \textbf{O2}    & \textbf{O3a}   & \textbf{O3b}   & \textbf{O4}     \\ \hline
\textbf{$\sigma_\mathrm{Z}=0.2$} & 18.2                                                                                    & 22.3$\pm5.4$  & 59.8$\pm8.7$   & 148.5$\pm10.2$ & 159.8$\pm13.0$ & 439.0$\pm26.3$  \\
\textbf{$\sigma_\mathrm{Z}=0.3$} & 40.0                                                                                    & 33.6$\pm8.3$  & 100.9$\pm11.1$ & 285.2$\pm22.9$ & 297.2$\pm20.4$ & 893.6$\pm38.5$  \\
\textbf{$\sigma_\mathrm{Z}=0.4$} & 100.4                                                                                   & 74.2$\pm13.1$ & 221.5$\pm13.0$ & 640.2$\pm31.6$ & 676.9$\pm32.7$ & 1995.9$\pm48.0$ \\ \hline
\textbf{LVK}                     & 23.9$^{+14.9}_{-8.6}$                                                                   & 15.5          & 17.3           & 99.7           & 69.5           & --              \\ \hline
\end{tabular}
\caption{Number of detected events from the catalogs. For comparison, $\mathcal{R}_0$ represents the local rate given in \cite{Mapelli_2021}. The errors are provided with respect to statistical fluctuations in the inclination and position of the source.}
\label{tab:detections}
\end{table}

\subsection{Predictions for 3G detectors}
\label{predic_3G}

We extend our study to the design of future ground-based detectors, such as ET, which will consist of a set of six interferometers with 10 km arms configured in a triangular arrangement, and two CE detectors with 40 km arms aligned similarly to the current Hanford and Livingston detectors.

\subsubsection{Individual events}
\label{sec:predic_3G_indi}

Table \ref{tab:ndet_3G} presents the number of detected BBH mergers from our astrophysical model, along with the associated errors from statistical fluctuations over inclination and sky position.

\begin{table}[]
\centering
\begin{tabular}{l|c|c|c}
\cline{1-4}
                                      & \textbf{$\sigma_Z=0.2$} & \textbf{$\sigma_Z=0.3$} & \textbf{$\sigma_Z=0.4$} \\ \hline
\multicolumn{1}{l|}{\textbf{ET}}     & 61570 (76.2\%)          & 93545 (79.0\%)          & 145867 (81.5\%)         \\
\multicolumn{1}{l|}{\textbf{2CE}}    & 78598 (97.3\%)          & 115714 (97.7\%)         & 175428 (98.0\%)         \\
\multicolumn{1}{l|}{\textbf{ET+2CE}} & 80074 (99.1\%)          & 117596 (99.3\%)         & 177906 (99.5\%)          \\ \hline
\multicolumn{1}{l|}{\textbf{Total}}  & 80762                   & 118429                  & 178883                  \\ \hline
\end{tabular}
\caption{Detection of BBH mergers in 3G detectors for the three models, assuming perfect matched filtering.}
\label{tab:ndet_3G}
\end{table}

\begin{figure}[h!]
    \centering
    \includegraphics[width=8cm]{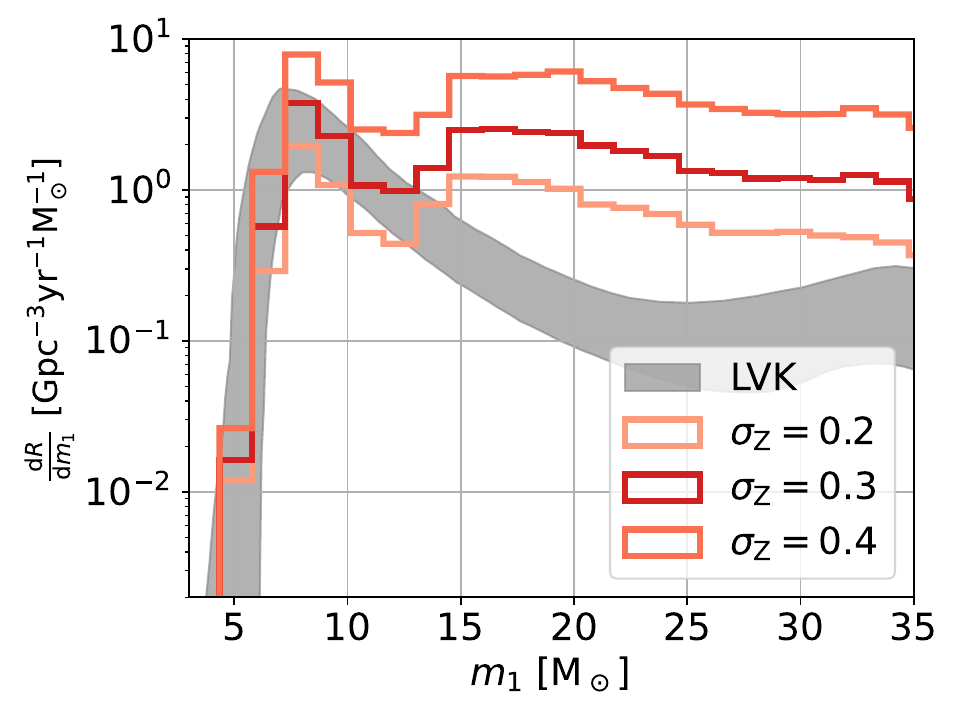}
    \caption{Distribution of m1 the most massive object of the binary for the three $\sigma_\mathrm{Z}$ variations. The grey area stands for the las LVK results \cite{LVK_gwtc3_population_2021}}
    \label{fig:Mc_distrib}
\end{figure}

For each model, the 2CE network detects more sources than ET, approximately 20\% more. The addition of ET to the 2CE network allows the future 3G configuration to capture nearly all sources from the catalog, missing at most $<1\%$ in the case of $\sigma_\mathrm{Z} = 0.2$. Across the different models, the number of detected events remains relatively consistent, although we note that the proportion of detected events increases slightly with $\sigma_\mathrm{Z}$. This trend is attributed to the primary mass $m_1$ distributions; as shown in Figure \ref{fig:Mc_distrib}, higher $\sigma_\mathrm{Z}$ values correlate with a greater proportion of events possessing large primary masses, making them more detectable.

\subsubsection{Astrophysical background prediction}
\label{sec:predic_3G_bkg}

In addition to the individual events the future 3G detectors expect to resolve an astrophysical background linked to the compact object population. 

\begin{figure}[h!]
    \centering
    \includegraphics[width=15cm]{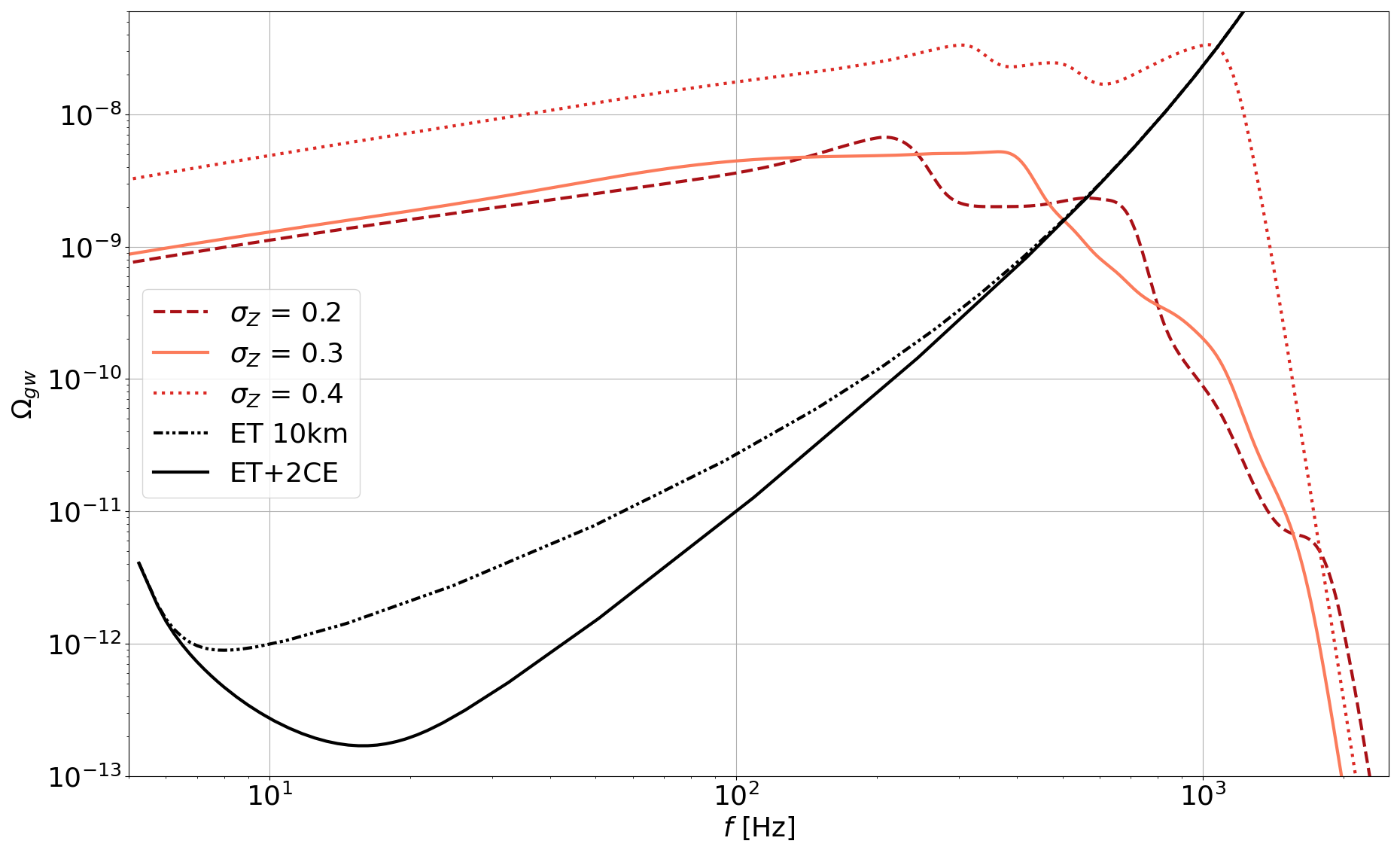}
    \caption{Background spectrum for the three catalogs in the study. The black lines represent the PICs of the third-generation networks.}
    \label{fig:Omega}
\end{figure}

Figure~\ref{fig:Omega} illustrates the total background computed for the three models used previously: $\sigma_\mathrm{Z}=0.2$, $\sigma_\mathrm{Z}=0.3$, and $\sigma_\mathrm{Z}=0.4$. The $f^{2/3}$ portion of the spectrum up to frequencies of $\sim 100$ Hz, $\sim 200$ Hz, and $\sim 300$ Hz for the models $\sigma_\mathrm{Z}=0.2$, $\sigma_\mathrm{Z}=0.3$, and $\sigma_\mathrm{Z}=0.4$, respectively, represents the typical shape arising from the inspiral phase of the coalescence. Above these frequencies, sources merge and do not emit gravitational waves, which explains the decrease in all model spectra. The different models exhibit varying spectral shapes due to differences in chirp masses and redshift distributions, as shown in \cite{Perigois_2020, Martinovic_2021, Perigois_2022}. For comparison, the figure also displays the Power Integrated Curves (PICs) of the networks studied, as defined in \cite{Thrane_2013}; a PIC tangent to a background spectrum signifies a background detectability SNR of 2 after one year of effective observation time.

Table \ref{tab:snr} shows the expected SNR for different detector configurations, for both the total and residual background. As we expect the total background is loud and will be clearly detecable.
However once we subtract the resolved event with a $\rho>8$ the background amplitude drastically decreases, up the undetectability in the case of ET+2CE with an SNR$<$1. 

\begin{table}[]
\centering
\begin{tabular}{ll|ccc}
\hline
                                                             &          & ET   & 2CE   & ET+2CE \\ \hline
\multicolumn{1}{l|}{\multirow{2}{*}{$\sigma_{\rm z} = 0.2$}} & Total    & 1701 & 3086  & 3720   \\
\multicolumn{1}{l|}{}                                        & Residual & 59   & 5     & $<$1   \\ \hline
\multicolumn{1}{l|}{\multirow{2}{*}{$\sigma_{\rm z} = 0.3$}} & Total    & 1969 & 3601  & 4332   \\
\multicolumn{1}{l|}{}                                        & Residual & 88   & 8     & $<$1   \\ \hline
\multicolumn{1}{l|}{\multirow{2}{*}{$\sigma_{\rm z} = 0.4$}} & Total    & 5739 & 10672 & 12786  \\
\multicolumn{1}{l|}{}                                        & Residual & 137  & 11    & $<$1   \\ \hline
\end{tabular}
\caption{Background SNRs for designed third-generation detectors.}
\label{tab:snr}
\end{table}

\section{Conclusion}
\label{sec:conclusion}

The Princess project proposes a tool to predict gravitational wave detections from compact binary mergers in ground-based detectors. It forecasts the observation of both individual signals and the background signal. In this paper, we presented the code along with a small example of its application.

We compute the predictions of gravitational wave detections based on three astrophysical models, assuming different metallicity spreads around the metallicity-redshift relation, with standard deviations of $\sigma_Z = 0.2$, $0.3$, or $0.4$. These three populations are mixtures of four different environments for the formation and evolution of BBHs, namely: field, young star clusters, nuclear star clusters, and globular clusters, as referenced in \cite{Mapelli_2021}.

The program incorporates the efficiency of each run to account for the actual time the detectors are in science mode, and evaluates the uncertainties arising from a random distribution of mergers in the sky and random inclinations. Despite strong efforts to model and quantify the quality of each run, the \princess{} code tends to overestimate the number of individual events detected by a factor of 2 to 10. This overestimation stems from the merger rates of the models, which are higher than the latest results from LIGO-Virgo-KAGRA \cite{LVK_gwtc3_population_2021}. Additionally, the uncertainties derived from the positions in the sky and the inclinations of the sources yield large error bars, making all results potentially consistent with previous science runs. We predict 400-500 individual detections of BBHs for the O4 observing run, assuming the model $\sigma_Z = 0.2$, which aligns most closely with the latest LVK results \cite{LVK_gwtc3_population_2021}. Based on this model, we forecast the detection rates for future ground-based detectors, including the Einstein Telescope (ET), a pair of Cosmic Explorers (CE), and a network comprising ET and 2CE. We show that this rate is biased by the mass distributions; higher masses lead to a higher detection rate. We predict a resolution of approximately 80\% for ET, 97-98\% for the pair of CEs, and greater than 99\% in the case of a network combining ET and 2CE.

Using this model, we also compute the expected total background spectrum. In agrement with previous studies \cite{Perigois_2020, Perigois_2022, Bavera_2021, Bellie_2023, Renzini_2024}, we demonstrate that a spectrum formed by the superposition of compact binary mergers would be widely detected with an SNR greater than 1700. Assuming we can subtract the individually resolved mergers with an SNR greater than 8 from the data, we could reduce the background by factors yet to be determined for ET, 2CE, and ET+2CE. This so-called residual background would still exhibit significant SNR in the cases of ET and 2CE but would yield an insignificant SNR (less than 3) in the network case of ET+2CE.

\subsection*{Future improvements}

The \princess{} tool aims to undergo upgrades in the near future. Our primary goals for the program include extending its capabilities to detect more massive BBHs and consequently expanding the observational range to lower frequencies. In this regard, we aim to incorporate new detector projects such as LISA and TianQuin, as well as pulsar timing array (PTA) estimations. From an astrophysical perspective, it would be beneficial to account for binary eccentricity in waveform computations and to include additional tools for new sources, such as extreme mass ratio inspirals (EMRIs) and asymmetric neutron stars.

In addition to these improvements, \princess{} would benefit from enhancements for third-generation detectors. We are currently adapting methods used in LVK for 3G detectors. While this yields promising estimates, further refinements are possible through more accurate tools, such as incorporating Earth rotation into waveform computations and upgrading individual SNR calculations to account for potential source superpositions.

\subsection*{Join the project}

\princess{} is an open project, and contributions for improvements, as well as new dependencies to existing codes, are welcome. Do not hesitate to contact the project lead to get organized. If users have any feedback, comments, or proposals for improvements, the project team is eager and open to discussion.

\subsection*{Acknowledgements}

CP thanks Florian Aubin, and Filippo Santoliquido for their constructive discussions and valuable advices. CP also expresses gratitude to the DEMOBLACK group for their feedback. Financial support from the European Research Council is acknowledged for the ERC Consolidator grant DEMOBLACK under contract no. 770017.
CP thanks the PnP referee for their valuable comments and insightful remarks. \\
This research has made use of data or software obtained from the Gravitational Wave Open Science Center (gwosc.org), a service of the LIGO Scientific Collaboration, the Virgo Collaboration, and KAGRA. This material is based upon work supported by NSF's LIGO Laboratory which is a major facility fully funded by the National Science Foundation, as well as the Science and Technology Facilities Council (STFC) of the United Kingdom, the Max-Planck-Society (MPS), and the State of Niedersachsen/Germany for support of the construction of Advanced LIGO and construction and operation of the GEO600 detector. Additional support for Advanced LIGO was provided by the Australian Research Council. Virgo is funded, through the European Gravitational Observatory (EGO), by the French Centre National de Recherche Scientifique (CNRS), the Italian Istituto Nazionale di Fisica Nucleare (INFN) and the Dutch Nikhef, with contributions by institutions from Belgium, Germany, Greece, Hungary, Ireland, Japan, Monaco, Poland, Portugal, Spain. KAGRA is supported by Ministry of Education, Culture, Sports, Science and Technology (MEXT), Japan Society for the Promotion of Science (JSPS) in Japan; National Research Foundation (NRF) and Ministry of Science and ICT (MSIT) in Korea; Academia Sinica (AS) and National Science and Technology Council (NSTC) in Taiwan

\bibliographystyle{unsrt}
\bibliography{biblio}

\begin{thebibliography}{10}

\bibitem{LIGO_AdLIGO_2015}
J.~{Aasi}, B.~P. {Abbott}, R.~{Abbott}, T.~{Abbott}, M.~R. {Abernathy}, K.~{Ackley}, C.~{Adams}, T.~{Adams}, P.~{Addesso}, R.~X. {Adhikari}, V.~{Adya}, and C.~et~al. {Affeldt}.
\newblock {Advanced LIGO}.
\newblock {\em Classical and Quantum Gravity}, 32(7):074001, April 2015.

\bibitem{Virgo_adVirgo_2015}
F.~{Acernese}, M.~{Agathos}, K.~{Agatsuma}, D.~{Aisa}, N.~{Allemandou}, A.~{Allocca}, J.~{Amarni}, P.~{Astone}, G.~{Balestri}, G.~{Ballardin}, F.~{Barone}, and et~al.
\newblock {Advanced Virgo: a second-generation interferometric gravitational wave detector}.
\newblock {\em Classical and Quantum Gravity}, 32(2):024001, January 2015.

\bibitem{LIGO_2016a}
B.~P. Abbott et~al.
\newblock {Observation of Gravitational Waves from a Binary Black Hole Merger}.
\newblock {\em Phys. Rev. Lett.}, 116(6):061102, 2016.

\bibitem{LIGO_GW150914}
B.~P. Abbott et~al.
\newblock {Properties of the Binary Black Hole Merger GW150914}.
\newblock {\em Phys. Rev. Lett.}, 116(24):241102, 2016.

\bibitem{LIGO_2016_GR}
B.~P. Abbott et~al.
\newblock {Tests of general relativity with GW150914}.
\newblock {\em Phys. Rev. Lett.}, 116(22):221101, 2016.
\newblock [Erratum: Phys.Rev.Lett. 121, 129902 (2018)].

\bibitem{LIGO_2016b}
B.~P. Abbott et~al.
\newblock {Astrophysical Implications of the Binary Black-Hole Merger GW150914}.
\newblock {\em Astrophys. J. Lett.}, 818(2):L22, 2016.

\bibitem{LIGO_2016_O1}
B.~P. Abbott et~al.
\newblock {Binary Black Hole Mergers in the first Advanced LIGO Observing Run}.
\newblock {\em Phys. Rev. X}, 6(4):041015, 2016.
\newblock [Erratum: Phys.Rev.X 8, 039903 (2018)].

\bibitem{LVK_gwtc1_2019}
B.~P. {Abbott}, R.~{Abbott}, T.~D. {Abbott}, S.~{Abraham}, F.~{Acernese}, K.~{Ackley}, C.~{Adams}, R.~X. {Adhikari}, V.~B. {Adya}, C.~{Affeldt}, M.~{Agathos}, K.~{Agatsuma}, N.~{Aggarwal}, O.~D. {Aguiar}, L.~{Aiello}, A.~{Ain}, P.~{Ajith}, G.~{Allen}, A.~{Allocca}, M.~A. {Aloy}, and et~al.
\newblock {GWTC-1: A Gravitational-Wave Transient Catalog of Compact Binary Mergers Observed by LIGO and Virgo during the First and Second Observing Runs}.
\newblock {\em Physical Review X}, 9(3):031040, July 2019.

\bibitem{LVK_gwtc2_2021}
R.~{Abbott}, T.~D. {Abbott}, S.~{Abraham}, F.~{Acernese}, K.~{Ackley}, A.~{Adams}, C.~{Adams}, R.~X. {Adhikari}, V.~B. {Adya}, C.~{Affeldt}, M.~{Agathos}, K.~{Agatsuma}, N.~{Aggarwal}, O.~D. {Aguiar}, L.~{Aiello}, A.~{Ain}, P.~{Ajith}, S.~{Akcay}, G.~{Allen}, and et~al.
\newblock {GWTC-2: Compact Binary Coalescences Observed by LIGO and Virgo during the First Half of the Third Observing Run}.
\newblock {\em Physical Review X}, 11(2):021053, April 2021.

\bibitem{LVK_gwtc21_2021}
R.~{Abbott}, T.~D. {Abbott}, F.~{Acernese}, K.~{Ackley}, C.~{Adams}, N.~{Adhikari}, R.~X. {Adhikari}, V.~B. {Adya}, C.~{Affeldt}, D.~{Agarwal}, M.~{Agathos}, K.~{Agatsuma}, N.~{Aggarwal}, O.~D. {Aguiar}, L.~{Aiello}, A.~{Ain}, and et~al.
\newblock {GWTC-2.1: Deep Extended Catalog of Compact Binary Coalescences Observed by LIGO and Virgo During the First Half of the Third Observing Run}.
\newblock {\em arXiv e-prints}, page arXiv:2108.01045, August 2021.

\bibitem{LVK_gwtc3_catalogue_2021}
R.~{Abbott}, T.~D. {Abbott}, F.~{Acernese}, K.~{Ackley}, C.~{Adams}, N.~{Adhikari}, R.~X. {Adhikari}, V.~B. {Adya}, C.~{Affeldt}, D.~{Agarwal}, M.~{Agathos}, K.~{Agatsuma}, N.~{Aggarwal}, O.~D. {Aguiar}, L.~{Aiello}, A.~{Ain}, P.~{Ajith}, S.~{Akcay}, and et~al.
\newblock {GWTC-3: Compact Binary Coalescences Observed by LIGO and Virgo During the Second Part of the Third Observing Run}.
\newblock {\em arXiv e-prints}, page arXiv:2111.03606, November 2021.

\bibitem{LIGO_GW170817}
B.~P. Abbott et~al.
\newblock {GW170817: Observation of Gravitational Waves from a Binary Neutron Star Inspiral}.
\newblock {\em Phys. Rev. Lett.}, 119(16):161101, 2017.

\bibitem{LIGO_GW190425}
B.~P. Abbott et~al.
\newblock {GW190425: Observation of a Compact Binary Coalescence with Total Mass $\sim 3.4 M_{\odot}$}.
\newblock {\em Astrophys. J. Lett.}, 892(1):L3, 2020.

\bibitem{LIGO_NSBH_2021}
R.~Abbott et~al.
\newblock {Observation of Gravitational Waves from Two Neutron Star\textendash{}Black Hole Coalescences}.
\newblock {\em Astrophys. J. Lett.}, 915(1):L5, 2021.

\bibitem{LVK_gwtc3_population_2021}
R.~Abbott et~al.
\newblock {Population of Merging Compact Binaries Inferred Using Gravitational Waves through GWTC-3}.
\newblock {\em Phys. Rev. X}, 13(1):011048, 2023.

\bibitem{KAGRA_2018}
T.~Akutsu et~al.
\newblock {KAGRA: 2.5 Generation Interferometric Gravitational Wave Detector}.
\newblock {\em Nature Astron.}, 3(1):35--40, 2019.

\bibitem{Mapelli_2021}
Michela Mapelli et~al.
\newblock {Hierarchical black hole mergers in young, globular and nuclear star clusters: the effect of metallicity, spin and cluster properties}.
\newblock {\em Mon. Not. Roy. Astron. Soc.}, 505(1):339--358, 2021.

\bibitem{Dominik_2012}
Michal Dominik, Krzysztof Belczynski, Christopher Fryer, Daniel Holz, Emanuele Berti, Tomasz Bulik, Ilya Mandel, and Richard O'Shaughnessy.
\newblock {Double Compact Objects I: The Significance of the Common Envelope on Merger Rates}.
\newblock {\em Astrophys. J.}, 759:52, 2012.

\bibitem{Rodriguez_2016}
Carl~L. Rodriguez, Sourav Chatterjee, and Frederic~A. Rasio.
\newblock {Binary Black Hole Mergers from Globular Clusters: Masses, Merger Rates, and the Impact of Stellar Evolution}.
\newblock {\em Phys. Rev. D}, 93(8):084029, 2016.

\bibitem{Coba}
Marica Branchesi et~al.
\newblock {Science with the Einstein Telescope: a comparison of different designs}.
\newblock {\em JCAP}, 07:068, 2023.

\bibitem{CE}
Matthew Evans et~al.
\newblock {A Horizon Study for Cosmic Explorer: Science, Observatories, and Community}.
\newblock 9 2021.

\bibitem{ET}
M.~Punturo et~al.
\newblock {The Einstein Telescope: A third-generation gravitational wave observatory}.
\newblock {\em Class. Quant. Grav.}, 27:194002, 2010.

\bibitem{Wang_2022}
Long Wang, Ataru Tanikawa, and Michiko Fujii.
\newblock {Gravitational wave of intermediate-mass black holes in Population III star clusters}.
\newblock {\em Mon. Not. Roy. Astron. Soc.}, 515(4):5106--5120, 2022.

\bibitem{Santoliquido_2023}
Filippo Santoliquido, Michela Mapelli, Giuliano Iorio, Guglielmo Costa, Simon C.~O. Glover, Tilman Hartwig, Ralf~S. Klessen, and Lorenzo Merli.
\newblock {Binary black hole mergers from Population III stars: uncertainties from star formation and binary star properties}.
\newblock {\em Mon. Not. Roy. Astron. Soc.}, 524(1):307--324, 2023.
\newblock [Erratum: Mon.Not.Roy.Astron.Soc. 528, 954--962 (2024)].

\bibitem{Amaro-Seoane_2009}
Pau Amaro-Seoane and Lucia Santamaria.
\newblock {Detection of IMBHs with ground-based gravitational wave observatories: A biography of a binary of black holes, from birth to death}.
\newblock {\em Astrophys. J.}, 722:1197--1206, 2010.

\bibitem{Jani_2019}
Karan Jani, Deirdre Shoemaker, and Curt Cutler.
\newblock {Detectability of Intermediate-Mass Black Holes in Multiband Gravitational Wave Astronomy}.
\newblock {\em Nature Astron.}, 4(3):260--265, 2019.

\bibitem{Liu_2024}
Boyuan Liu, Tilman Hartwig, Nina~S. Sartorio, Irina Dvorkin, Guglielmo Costa, Filippo Santoliquido, Anastasia Fialkov, Ralf~S. Klessen, and Volker Bromm.
\newblock {Gravitational waves from mergers of Population III binary black holes: roles played by two evolution channels}.
\newblock {\em Mon. Not. Roy. Astron. Soc.}, 534(3):1634--1667, 2024.

\bibitem{Mestichelli_2024}
Benedetta Mestichelli, Michela Mapelli, Stefano Torniamenti, Manuel~Arca Sedda, Marica Branchesi, Guglielmo Costa, Giuliano Iorio, and Filippo Santoliquido.
\newblock {Binary black hole mergers from Population III star clusters}.
\newblock 5 2024.

\bibitem{Vaccaro_2023}
Maria~Paola Vaccaro, Michela Mapelli, Carole P\'erigois, Dario Barone, Maria~Celeste Artale, Marco Dall'Amico, Giuliano Iorio, and Stefano Torniamenti.
\newblock {Impact of gas hardening on the population properties of hierarchical black hole mergers in active galactic nucleus disks}.
\newblock {\em Astron. Astrophys.}, 685:A51, 2024.

\bibitem{Belczynski_2016}
K.~Belczynski et~al.
\newblock {The Effect of Pair-Instability Mass Loss on Black Hole Mergers}.
\newblock {\em Astron. Astrophys.}, 594:A97, 2016.

\bibitem{Mapelli_2019}
Michela Mapelli, Nicola Giacobbo, Filippo Santoliquido, and M.~Celeste Artale.
\newblock {The properties of merging black holes and neutron stars across cosmic time}.
\newblock {\em Mon. Not. Roy. Astron. Soc.}, 487(1):2--13, 2019.

\bibitem{Giacobbo_2018}
Nicola Giacobbo and Michela Mapelli.
\newblock {The progenitors of compact-object binaries: impact of metallicity, common envelope and natal kicks}.
\newblock {\em Mon. Not. Roy. Astron. Soc.}, 480(2):2011--2030, 2018.

\bibitem{Woosley_2016}
S.~E. Woosley.
\newblock {Pulsational Pair-Instability Supernovae}.
\newblock {\em Astrophys. J.}, 836(2):244, 2017.

\bibitem{Kroupa_2021}
Pavel {Kroupa} and Tereza {Jerabkova}.
\newblock {The initial mass function of stars and the star-formation rates of galaxies}.
\newblock {\em arXiv e-prints}, page arXiv:2112.10788, December 2021.

\bibitem{Sana_2012}
H.~{Sana}, S.~E. {de Mink}, A.~{de Koter}, N.~{Langer}, C.~J. {Evans}, M.~{Gieles}, E.~{Gosset}, R.~G. {Izzard}, J.~B. {Le Bouquin}, and F.~R.~N. {Schneider}.
\newblock {Binary Interaction Dominates the Evolution of Massive Stars}.
\newblock {\em Science}, 337(6093):444, July 2012.

\bibitem{Bavera_2022}
Simone~S. Bavera et~al.
\newblock {The formation of $30\,M_\odot$ merging black holes at solar metallicity}.
\newblock 12 2022.

\bibitem{Annala_2017}
Eemeli Annala, Tyler Gorda, Aleksi Kurkela, and Aleksi Vuorinen.
\newblock {Gravitational-wave constraints on the neutron-star-matter Equation of State}.
\newblock {\em Phys. Rev. Lett.}, 120(17):172703, 2018.

\bibitem{Bavera_2021}
Simone~S. Bavera, Gabriele Franciolini, Giulia Cusin, Antonio Riotto, Michael Zevin, and Tassos Fragos.
\newblock {Stochastic gravitational-wave background as a tool for investigating multi-channel astrophysical and primordial black-hole mergers}.
\newblock {\em Astron. Astrophys.}, 660:A26, 2022.

\bibitem{Perigois_2020}
C.~P\'erigois, C.~Belczynski, T.~Bulik, and T.~Regimbau.
\newblock {StarTrack predictions of the stochastic gravitational-wave background from compact binary mergers}.
\newblock {\em Phys. Rev. D}, 103(4):043002, 2021.

\bibitem{Perigois_2022}
Carole P\'erigois, Filippo Santoliquido, Yann Bouffanais, Ugo~N. Di~Carlo, Nicola Giacobbo, Sara Rastello, Michela Mapelli, and Tania Regimbau.
\newblock {Gravitational background from dynamical binaries and detectability with 2G detectors}.
\newblock {\em Phys. Rev. D}, 105(10):103032, 2022.

\bibitem{Pycbc-2005}
Bruce Allen, Warren~G. Anderson, Patrick~R. Brady, Duncan~A. Brown, and Jolien D.~E. Creighton.
\newblock {FINDCHIRP: An Algorithm for detection of gravitational waves from inspiraling compact binaries}.
\newblock {\em Phys. Rev. D}, 85:122006, 2012.

\bibitem{Pycbc-2014}
Tito Dal~Canton et~al.
\newblock {Implementing a search for aligned-spin neutron star-black hole systems with advanced ground based gravitational wave detectors}.
\newblock {\em Phys. Rev. D}, 90(8):082004, 2014.

\bibitem{Pycbc-2017}
Alexander~H. Nitz, Thomas Dent, Tito Dal~Canton, Stephen Fairhurst, and Duncan~A. Brown.
\newblock {Detecting binary compact-object mergers with gravitational waves: Understanding and Improving the sensitivity of the PyCBC search}.
\newblock {\em Astrophys. J.}, 849(2):118, 2017.

\bibitem{Pandas}
Wes McKinney et~al.
\newblock Data structures for statistical computing in python.
\newblock In {\em Proceedings of the 9th Python in Science Conference}, volume 445, pages 51--56. Austin, TX, 2010.

\bibitem{Numpy}
Charles~R. Harris, K.~Jarrod Millman, St{\'{e}}fan~J. van~der Walt, Ralf Gommers, Pauli Virtanen, David Cournapeau, Eric Wieser, Julian Taylor, Sebastian Berg, Nathaniel~J. Smith, Robert Kern, Matti Picus, Stephan Hoyer, Marten~H. van Kerkwijk, Matthew Brett, Allan Haldane, Jaime~Fern{\'{a}}ndez del R{\'{i}}o, Mark Wiebe, Pearu Peterson, Pierre G{\'{e}}rard-Marchant, Kevin Sheppard, Tyler Reddy, Warren Weckesser, Hameer Abbasi, Christoph Gohlke, and Travis~E. Oliphant.
\newblock Array programming with {NumPy}.
\newblock {\em Nature}, 585(7825):357--362, September 2020.

\bibitem{SciPy_2020}
Pauli Virtanen, Ralf Gommers, Travis~E. Oliphant, Matt Haberland, Tyler Reddy, David Cournapeau, Evgeni Burovski, Pearu Peterson, Warren Weckesser, Jonathan Bright, St{\'e}fan~J. {van der Walt}, Matthew Brett, Joshua Wilson, K.~Jarrod Millman, Nikolay Mayorov, Andrew R.~J. Nelson, Eric Jones, Robert Kern, Eric Larson, C~J Carey, {\.I}lhan Polat, Yu~Feng, Eric~W. Moore, Jake {VanderPlas}, Denis Laxalde, Josef Perktold, Robert Cimrman, Ian Henriksen, E.~A. Quintero, Charles~R. Harris, Anne~M. Archibald, Ant{\^o}nio~H. Ribeiro, Fabian Pedregosa, Paul {van Mulbregt}, and {SciPy 1.0 Contributors}.
\newblock {{SciPy} 1.0: Fundamental Algorithms for Scientific Computing in Python}.
\newblock {\em Nature Methods}, 17:261--272, 2020.

\bibitem{astropy}
{Astropy Collaboration} and {Astropy Project Contributors}.
\newblock {The Astropy Project: Sustaining and Growing a Community-oriented Open-source Project and the Latest Major Release (v5.0) of the Core Package}.
\newblock {\em ApJ}, 935(2):167, August 2022.

\bibitem{Ajith_2007}
P.~Ajith et~al.
\newblock {A Template bank for gravitational waveforms from coalescing binary black holes. I. Non-spinning binaries}.
\newblock {\em Phys. Rev. D}, 77:104017, 2008.
\newblock [Erratum: Phys.Rev.D 79, 129901 (2009)].

\bibitem{Ajith_2009}
P.~Ajith et~al.
\newblock {Inspiral-merger-ringdown waveforms for black-hole binaries with non-precessing spins}.
\newblock {\em Phys. Rev. Lett.}, 106:241101, 2011.

\bibitem{Gerosa_2017}
Davide Gerosa and Emanuele Berti.
\newblock {Are merging black holes born from stellar collapse or previous mergers?}
\newblock {\em Phys. Rev. D}, 95(12):124046, 2017.

\bibitem{Finn_1992}
Lee~S. Finn.
\newblock {Detection, measurement and gravitational radiation}.
\newblock {\em Phys. Rev. D}, 46:5236--5249, 1992.

\bibitem{Allen_1997}
Bruce Allen and Joseph~D. Romano.
\newblock {Detecting a stochastic background of gravitational radiation: Signal processing strategies and sensitivities}.
\newblock {\em Phys. Rev. D}, 59:102001, 1999.

\bibitem{Phinney_2001}
E.~S. Phinney.
\newblock {A Practical theorem on gravitational wave backgrounds}.
\newblock 7 2001.

\bibitem{Regimbau_2014}
Tania Regimbau, Duncan Meacher, and Michael Coughlin.
\newblock {Second Einstein Telescope mock science challenge: Detection of the gravitational-wave stochastic background from compact binary coalescences}.
\newblock {\em Phys. Rev. D}, 89(8):084046, 2014.

\bibitem{Flanagan_1993}
Eanna~E. Flanagan.
\newblock {The Sensitivity of the laser interferometer gravitational wave observatory (LIGO) to a stochastic background, and its dependence on the detector orientations}.
\newblock {\em Phys. Rev. D}, 48:2389--2407, 1993.

\bibitem{Christensen_1992}
Nelson {Christensen}.
\newblock {Measuring the stochastic gravitational-radiation background with laser-interferometric antennas}.
\newblock {\em Phys. Rev. D}, 46(12):5250--5266, December 1992.

\bibitem{GWOSC_2021}
Rich Abbott et~al.
\newblock {Open data from the first and second observing runs of Advanced LIGO and Advanced Virgo}.
\newblock {\em SoftwareX}, 13:100658, 2021.

\bibitem{GWOSC_2023}
R.~Abbott et~al.
\newblock {Open Data from the Third Observing Run of LIGO, Virgo, KAGRA, and GEO}.
\newblock {\em Astrophys. J. Suppl.}, 267(2):29, 2023.

\bibitem{Broekgaarden_2021}
Floor~S. Broekgaarden et~al.
\newblock {Impact of Massive Binary Star and Cosmic Evolution on Gravitational Wave Observations II: Double Compact Object Rates and Properties}.
\newblock 12 2021.

\bibitem{Martinovic_2021}
Katarina Martinovic, Carole Perigois, Tania Regimbau, and Mairi Sakellariadou.
\newblock {Footprints of Population III Stars in the Gravitational-wave Background}.
\newblock {\em Astrophys. J.}, 940(1):29, 2022.

\bibitem{Thrane_2013}
Eric Thrane and Joseph~D. Romano.
\newblock {Sensitivity curves for searches for gravitational-wave backgrounds}.
\newblock {\em Phys. Rev. D}, 88(12):124032, 2013.

\bibitem{Bellie_2023}
Darsan~S. Bellie, Sharan Banagiri, Zoheyr Doctor, and Vicky Kalogera.
\newblock {Unresolved stochastic background from compact binary mergers detectable by next-generation ground-based gravitational-wave observatories}.
\newblock {\em Phys. Rev. D}, 110(2):023006, 2024.

\bibitem{Renzini_2024}
Arianna~I. Renzini and Jacob Golomb.
\newblock {Projections of the uncertainty on the compact binary population background using popstock}.
\newblock {\em Astron. Astrophys.}, 691:A238, 2024.

\bibitem{LVK_O1_sensitivity}
Benjamin~P. Abbott et~al.
\newblock {Sensitivity of the Advanced LIGO detectors at the beginning of gravitational wave astronomy}.
\newblock {\em Phys. Rev. D}, 93(11):112004, 2016.
\newblock [Addendum: Phys.Rev.D 97, 059901 (2018)].

\bibitem{ET_science_2023}
Marica Branchesi et~al.
\newblock {Science with the Einstein Telescope: a comparison of different designs}.
\newblock {\em JCAP}, 07:068, 2023.

\bibitem{Evans_2021}
Matthew Evans et~al.
\newblock {A Horizon Study for Cosmic Explorer: Science, Observatories, and Community}.
\newblock 9 2021.

\bibitem{Srivastava_2022}
Varun Srivastava, Derek Davis, Kevin Kuns, Philippe Landry, Stefan Ballmer, Matthew Evans, Evan~D. Hall, Jocelyn Read, and B.~S. Sathyaprakash.
\newblock {Science-driven Tunable Design of Cosmic Explorer Detectors}.
\newblock {\em Astrophys. J.}, 931(1):22, 2022.

\end{thebibliography}

\appendix

\section{Requirement and quick start}
\label{app:quick_start}
\princess{} is a public code available on GitHub\footnote{\href{https://github.com/Cperigois/Princess}{https://github.com/Cperigois/Princess}} and GitLab\footnote{\href{https://gitlab.com/Cperigois/Princess}{https://gitlab.com/Cperigois/Princess}}. For a first used the directory can be simply cloned. The programm is requiering several package in specific version:
\begin{itemize}
	\item Python 3.7 or better
	\item PyCBC v1.18 (more recent version are \textbf{NOT} working with \princess)\cite{Pycbc-2005, Pycbc-2014, Pycbc-2017}
	\item Pandas \cite{Pandas}
	\item NumPy \cite{Numpy}
	\item SciPy \cite{SciPy_2020}
\end{itemize}

\paragraph{Getting started with \princess}
In order to be as user friendly as possible the code contain a Getting Started file (.py and .ipynb), detailing all the steps for a basic study.
By default this Getting started file is linked to a test catalogue. After the first run the user can compare the obtained results with the ones in \texttt{\textbackslash AuxiliaryFiles\textbackslash checkDefault\textbackslash}. The GitHub and GitLab pages contain the link to the recording tutorial session \footnote{First session of tutorial in planned in fall spring 2023.}. 

\section{Column names for the input catalogs}\label{sec:cat_columns}

\begin{table}[]
\centering

\begin{tabular}{|l|c|l|}
\hline
\textbf{Name} & \textbf{Symbol} & \textbf{Description}                                                                        \\ \hline
Mc            & $\mathcal{M}_c$ & Chirp mass in the source frame [\msun]                                                      \\ \hline
q             & $q$             & Mass ratio, $q<1$                                                                           \\ \hline
m1            & $m_1$           & Mass of the first component (m1$>$m2) in the source frame [\msun]                             \\ \hline
m2            & $m_2$           & Mass of the first component (m1$>$m2) in the source frame [\msun]                             \\ \hline
s1            & $s_1$           & Spin amplitude of the first component                                                       \\ \hline
s2            & $s_2$           & Spin amplitude of the second component                                                      \\ \hline
theta1        & $\theta_1$      & Angle between $\vec{s_1}$ and angular momentum of the binary $\vec{L}$                      \\ \hline
theta2        & $\theta_1$      & Angle between $\vec{s_2}$ and angular momentum of the binary $\vec{L}$                      \\ \hline
a0            & $a_0$           & Semi-major axis of the trajectory at the formation of the second compact object [R$_\odot$] \\ \hline
e0            & $e_0$           & Eccentricity of the binary at the formation of the second compact object.                   \\ \hline
inc           & $\iota$         & Inclination angle of the binary with respect to the detector frame.                         \\ \hline
z             & $z$             & Redshift.                                                                                   \\ \hline
Dl            & $d_\mathrm{L}$  & Luminosity distance of the binary [Mpc]                                                     \\ \hline
flag          & --              & Optional columns where the user can se an identifer.                                        \\ \hline
\end{tabular}
\caption{Table of all parameters possible in the input catalogues of the user. Note that this list is NOT exhaustiv, and aims to remind the standard labels which are used in \princess.}
\label{tab:param_label}
\end{table}

\section{Spin options in the programm}
\label{app:spin_options}
\begin{figure}[h!]
    \centering
    \includegraphics[width=10cm]{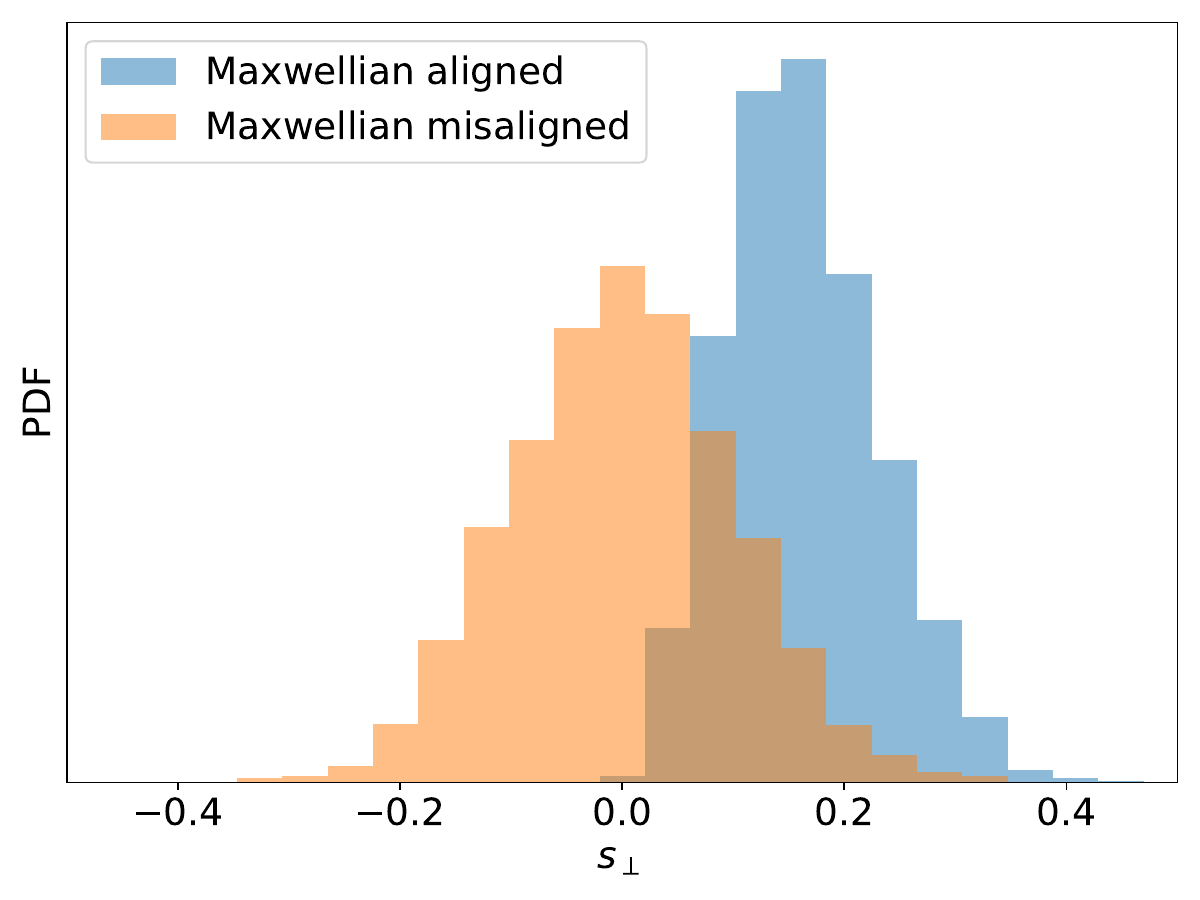}
    \caption{BBH detections predicted in Einstein Telescope by \princess{} for an astrophysical model including Isolated and Dynamical(young star clusters) evolutions. $f_\mathrm{Dyn}$ stands for the fraction of dynamical binaries in the total population.}
    \label{fig:spins}
\end{figure}

In this section are presented the different options avalable in \princess{} to set the spin of the binaries. Following the last obesrvations made by LIGO-Vigro \cite{LVK_gwtc3_population_2021}, spin amplitudes are drawn from Maxwellian laws with a parameter $a$=0.1. We then distinguish the case of isolated binaries where the spins are assumed to be aligned, and the case of dynamics where a random inclination of spins whith respect to the angular momentum of the binary is drawn randomly. Table \ref{tab:spins} summary the three options available yet. In addition Figure \ref{fig:spins} shows the distributions obtained from our different options.

\begin{itemize}
    \item \textbf{'In\_Catalogue'}: Spin projections are already included in the user catalogue.
    \item \textbf{'Chi\&Theta'}: Amplitude and theta angle are available in the catalogue. Thus, $s_{1(2)} = \chi_{1(2)} \cos\theta_{1(2)}$.
    \item \textbf{'Zeros'}: Assign a zero spin to all binaries.
    \item \textbf{'Rand'}: Generate random aligned spins.
    \item \textbf{'Rand\_Dynamics'}: Generate random spins with random orientations.
\end{itemize}

More details about how the spins are generated can be found in the appendix.

\section{Detectors available in the program}\label{sec:detection}

This section summerizes all PSDs available in \princess{} and some of the important ones used from PyCBC. 
In the table \ref{tab:PSD_princess} we show all files available in \texttt{/AuxiliaryFiles/PSDs} and how to call them while setting the detectors with the \textit{Detector class}. Depending on the runs the psds doe
\begin{table}[]
\centering
\caption{List of PSDs available in the Princess code. These can be call by setting \texttt{psd\_file} and the \texttt{origin = 'Princess'} when calling the detector class.}
\label{tab:PSD_princess}
\begin{tabular}{l|l|c|l}
\hline
\textbf{\begin{tabular}[c]{@{}l@{}}\texttt{psd\_file}\\ \\ name of the psd\end{tabular}} & \textbf{Detector}                                                                   & \textbf{Run}                                 & \textbf{Reference}                                                                                                                                                         \\ \hline
Livingston\_O1                                                                         & aLIGO, Livingston                                                                   & \multirow{2}{*}{O1}                          &             \cite{LVK_O1_sensitivity}                                                                                                                                                               \\
 Hanford\_O1                                                                            & aLIGO, Hanford                                                                      &                                              &           \cite{LVK_O1_sensitivity}                                                                                                                                                           \\ \hline
Livingston\_O2                                                                         & aLIGO, Livingston                                                                   & \multirow{3}{*}{O2}                          & \cite{LVK_gwtc1_2019}                                                                                                                                                        \\
Hanford\_O2                                                                            & aLIGO, Hanford                                                                      &                                              & \cite{LVK_gwtc1_2019}                                                                                                                                                        \\
Virgo\_O2                                                                              & adVirgo                                                                             &                                              & \cite{LVK_gwtc1_2019}                                                                                                                                                        \\ \hline
Livingston\_O3a                                                                        & aLIGO, Livingston                                                                   & \multirow{3}{*}{O3a}                         & \cite{LVK_gwtc2_2021}                                                                                                                                                        \\
Hanford\_O3a                                                                           & aLIGO, Hanford                                                                      &                                              & \cite{LVK_gwtc2_2021}                                                                                                                                                        \\
Virgo\_O3a                                                                             & adVirgo                                                                             &                                              & \cite{LVK_gwtc2_2021}                                                                                                                                                        \\ \hline
Livingston\_O3b                                                                        & aLIGO, Livingston                                                                   & \multirow{3}{*}{O3b}                         & \cite{LVK_gwtc21_2021}                                                                                                                                              \\
Hanford\_O3b                                                                           & aLIGO, Hanford                                                                      &                                              & \cite{LVK_gwtc21_2021}                                                                                                                                              \\
Virgo\_O3b                                                                             & adVirgo                                                                             &                                              & \cite{LVK_gwtc21_2021}                                                                                                                                              \\ \hline
Livingston\_O4                                                                         & aLIGO, Livingston                                                                   & \multirow{4}{*}{O4}                          & \cite{Pycbc-2017}                                                                                                                                                       \\
Hanford\_O4                                                                            & aLIGO, Hanford                                                                      &                                              & \cite{Pycbc-2017}                                                                                                                                                       \\
Virgo\_O4                                                                              & adVirgo                                                                             &                                              & \cite{Pycbc-2017}                                                                                                                                                       \\
Kagra\_O4                                                                              & KAGRA                                                                               &                                              & \cite{Pycbc-2017}                                                                                                                                                       \\ \hline
Livingston\_O5                                                                         & A+LIGO, Livingston                                                                  & \multirow{4}{*}{O5}                          & \cite{Pycbc-2017}                                                                                                                                                       \\
Hanford\_O5                                                                            & A+LIGO, Hanford                                                                     &                                              & \cite{Pycbc-2017}                                                                                                                                                       \\
Virgo\_O5                                                                              & AdVirgo +                                                                           &                                              & \cite{Pycbc-2017}                                                                                                                                                       \\
Kagra\_O5                                                                              & KAGRA                                                                               &                                              & \cite{Pycbc-2017}                                                                                                                                                       \\ \hline
aLIGO\_Des                                                                             & aLIGO                                                                               & \multirow{3}{*}{Design}                      &                                                                                                                                                                            \\
adVirgo\_Des                                                                           & AdVirgo                                                                             &                                              &                                                                                                                                                                            \\
Kagra\_Des                                                                             & KAGRA                                                                               &                                              &                                                                                                                                                                            \\ \hline
ET10\_CoBa                                                                             & \begin{tabular}[c]{@{}l@{}}Einstein Telescope,\\ 10km arms\end{tabular}             & \multicolumn{1}{l|}{\multirow{3}{*}{Design}} & \multirow{3}{*}{\begin{tabular}[c]{@{}l@{}}\cite{ET_science_2023}\\ Triangle shape assuming combination\\ of low-frequency and high frequency pipes.\\ (LF+HF)\end{tabular}} \\
ET15\_CoBa                                                                             & \begin{tabular}[c]{@{}l@{}}Einstein Telescope,\\ Triangle of 15km arms\end{tabular} & \multicolumn{1}{l|}{}                        &                                                                                                                                                                            \\
ET20\_CoBa                                                                             & \begin{tabular}[c]{@{}l@{}}Einstein Telescope,\\ Triangle of 20km arms\end{tabular} & \multicolumn{1}{l|}{}                        &                                                                                                                                                                            \\ \hline
CE\_20km                                                                               & \begin{tabular}[c]{@{}l@{}}Cosmic Explorer,\\ 20km arms\end{tabular}                & --                                           & \begin{tabular}[c]{@{}l@{}}\cite{Evans_2021,Srivastava_2022},\\ Tuned for post-merger signals\end{tabular}                                                                 \\
CE\_40km                                                                               & \begin{tabular}[c]{@{}l@{}}Cosmic Explorer,\\ 40km arms\end{tabular}                & --                                           & \begin{tabular}[c]{@{}l@{}}\cite{Evans_2021,Srivastava_2022},\\ Tuned for low-frequency signals\end{tabular}                                                               \\ \hline
\end{tabular}
\end{table}

\begin{figure}[h!]
    \centering
    \includegraphics[width=14cm]{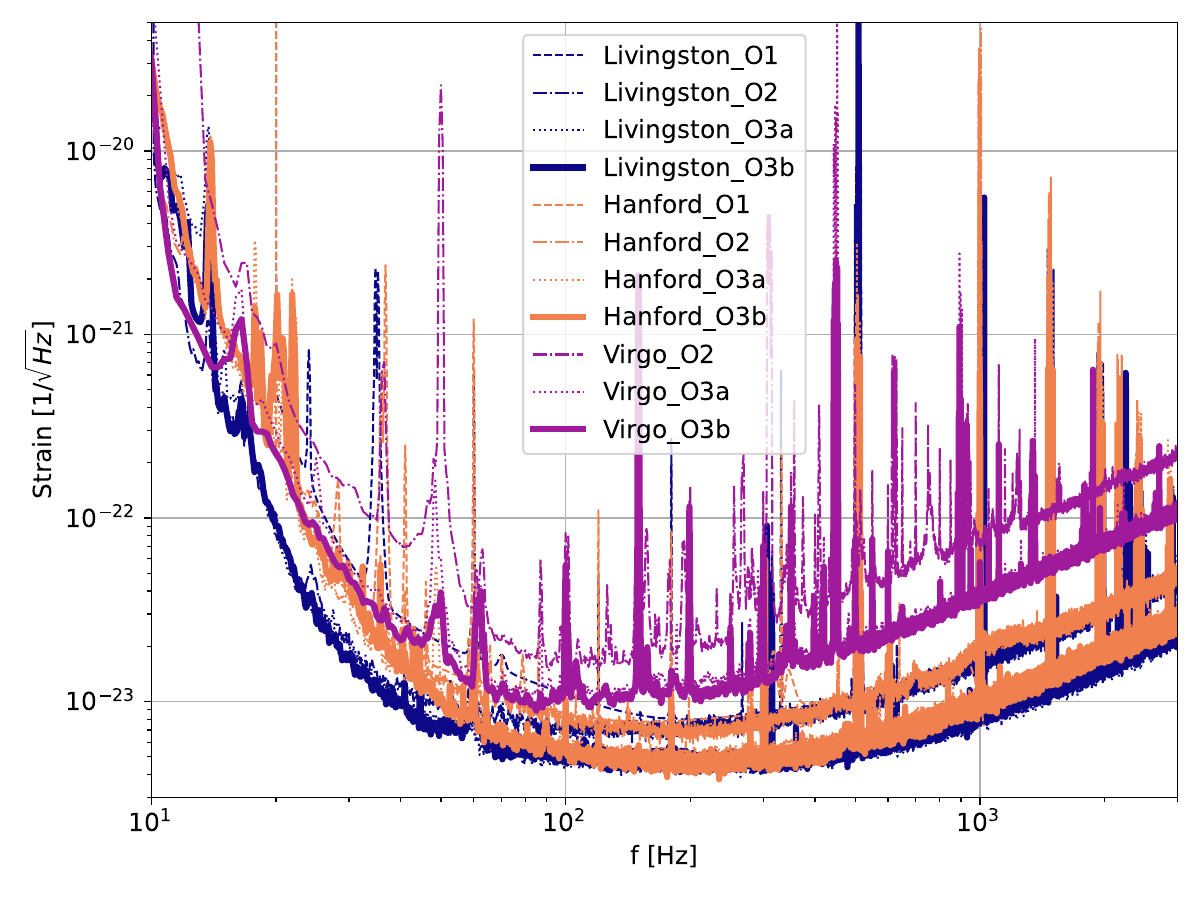}
    \caption{Sensitivities from previous runs available in \princess.}
    \label{fig:my_label}
\end{figure}

\begin{figure}[h!]
    \centering
    \includegraphics[width=14cm]{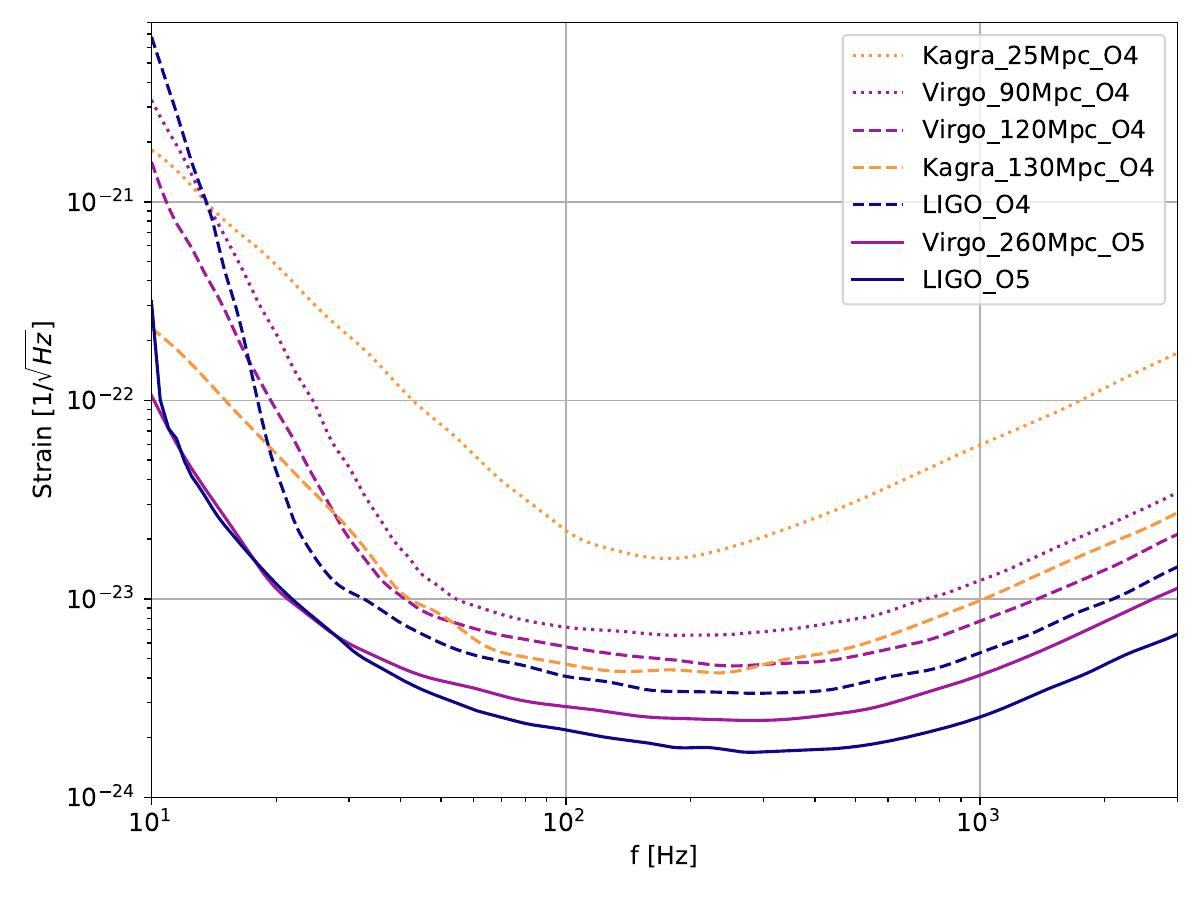}
    \caption{Sensitivities for future second generation runs \cite{Pycbc-2017} available in \princess.}
    \label{fig:my_label}
\end{figure}

\begin{table}[]
\centering
\begin{tabular}{|l|l|}
\hline
\textbf{Name} & \textbf{Description}                                                \\ \hline
H             & Detector located and oriented as LIGO-Hanford.                      \\ \hline
L             & Detector located and oriented as LIGO-Livingston.                   \\ \hline
V             & Detector located and oriented as Virgo.                             \\ \hline
ET            & Set of three detectors in triangle, with an arm opening of $\pi/3$. \\ \hline
\end{tabular}
\caption{Configurations available to set up detetctors.}
\label{tab:config_detectors}
\end{table}

\section{Advanced parameters}
\label{app:advparam}

The file \texttt{advanced\_params.py} contains all the parameter not needed by the user for a basic use. 

\dirtree{%
.1 Princess.
.2 AuxiliaryFiles.
.2 Run.
.3 \textbf{advanced\_params.py}.
.3 getting\_started.py.
.2 astrotools.
.2 gwtools.
.2 stochastic.
.2 README.md.
.2 run.py.
}

\begin{table}[]
\hspace{-1.1cm}
\begin{tabular}{ll|l}
\hline
\multicolumn{2}{l|}{\textbf{Variable}}                                     & \textbf{Description}                                                                        \\ \hline
\multicolumn{2}{l|}{\texttt{types}}                                        & \begin{tabular}[c]{@{}l@{}}Dictionary containing parameters for different types\\ of detectors among {2G, 3G, LISA, PTA}\end{tabular} \\ \hline
\multicolumn{1}{l|}{}        & \texttt{freq}                               & Minimum and maximum frequencies for each detector                                           \\
\multicolumn{1}{l|}{}        & \texttt{ref}                                & Reference frequency used to evaluate the background. \ref{eq:ratio}                         \\
\multicolumn{1}{l|}{}        & \texttt{waveform}                           & Gravitational wave approximant used                                                         \\ \hline
\multicolumn{2}{l|}{\texttt{input\_parameters}}                            & User input catalogue parameters                                                             \\ \hline
\multicolumn{1}{l|}{}        & \texttt{m1}, \texttt{m2}                    & Masses of compact objects 1 and 2                                                           \\
\multicolumn{1}{l|}{}        & \texttt{chi1}, \texttt{chi2}                & Spin magnitudes of compact objects 1 and 2                                                  \\
\multicolumn{1}{l|}{}        & \texttt{th1}, \texttt{th2}                  & Angles between angular momentum and spin                                                    \\
\multicolumn{1}{l|}{}        & \texttt{cos\_nu1}, \texttt{cos\_nu2}        & Cosine of tilt angles during supernovae                                                     \\
\multicolumn{1}{l|}{}        & \texttt{z\_merg}, \texttt{z\_form}          & Redshift at merger and formation                                                            \\
\multicolumn{1}{l|}{}        & \texttt{time\_delay}                        & Time delay between events                                                                   \\ \hline
\multicolumn{2}{l|}{\texttt{sampling\_size}}                               & Sample size to be used                                                                      \\
\multicolumn{2}{l|}{\texttt{sampling\_number\_of\_walkers}}                & Number of MCMC walkers                                                                      \\
\multicolumn{2}{l|}{\texttt{sampling\_chain\_length}}                      & MCMC chain length                                                                           \\
\multicolumn{2}{l|}{\texttt{sampling\_bandwidth\_KDE}}                     & KDE bandwidth to use                                                                        \\ \hline
\multicolumn{2}{l|}{\texttt{detectors\_avail}}                             & List of available detectors                                                                 \\ \hline
\multicolumn{1}{l|}{}        & \texttt{psd\_attributes}                    & PSD parameters for each detector                                                            \\
\multicolumn{1}{l|}{}        & \texttt{psd\_name}                          & Name of the PSD file for each detector                                                      \\
\multicolumn{1}{l|}{}        & \texttt{min\_freq}, \texttt{max\_freq}      & Minimum and maximum frequencies for each PSD                                                \\
\multicolumn{1}{l|}{}        & \texttt{delta\_freq\_min}                   & Minimum delta frequency for each PSD                                                        \\ \hline
\multicolumn{2}{l|}{\texttt{pAstroLimit}}                                  & pAstro limit for selecting GW events                                                        \\
\multicolumn{2}{l|}{\texttt{farLimit}}                                     & FAR limit for GW events                                                                     \\
\multicolumn{2}{l|}{\texttt{snrLimit}}                                     & SNR limit for GW events                                                                     \\
\multicolumn{2}{l|}{\texttt{available\_obs\_runs}}                         & List of available observation runs                                                          \\ \hline
\multicolumn{2}{l|}{\texttt{bayes\_model\_processing\_waveform\_approximant}} & Waveform approximant used in the model                                                      \\
\multicolumn{2}{l|}{\texttt{option\_SNR\_computation}}                     & Option for SNR computation                                                                  \\
\multicolumn{2}{l|}{\texttt{bayes\_model\_processing\_bandwidth\_KDE}}     & KDE bandwidth for processing                                                                \\
\multicolumn{2}{l|}{\texttt{bayes\_option\_compute\_likelihood}}           & Option for likelihood computation                                                           \\
\multicolumn{2}{l|}{\texttt{bayes\_option\_multichannel}}                  & Multichannel option for data processing                                                     \\ \hline
\end{tabular}
\caption{Table encapsulating the main parameters from the file \texttt{advanced\_params.py}.}
\label{tab:my-table}
\end{table}

\section{Auxiliary files}
\label{app:auxfiles}
The following tree highlight the \texttt{\textbf{AuxiliaryFiles}} folder and its content.

\dirtree{%
.1 Princess.
.2 \textbf{AuxiliaryFiles}.
.3 \textbf{LVK\_data}.
.4 \textbf{Posterior}.
.5 \texttt{\textbf{<name of the event>\_post.dat}}.
.4 \textbf{Prior}.
.5 \texttt{\textbf{<name of the event>\_prior.dat}}.
.4 \textbf{GWlist.csv}.
.3 \textbf{ORFs}.
.4 \textbf{ORF.dat}.
.3 \textbf{PICs}.
.3 \textbf{PSDs}.
.3 \textbf{dl\_z\_table\_Planck\_15.txt}.
.3 \textbf{factor\_table.dat}.
.2 Run.
.3 advanced\_params.py.
.3 getting\_started.py.
.2 astrotools.
.2 gwtools.
.2 stochastic.
.2 README.md.
.2 run.py.
}

\begin{itemize}
    \item \texttt{\textbf{LVK\_data}} contains all the file related to the gravitational wave events detected by the LVK collaboration up to O3b. The two folders \texttt{\textbf{Posterior}} and \texttt{\textbf{Prior}} contain the posteriors and prior of all events. The file \texttt{\textbf{GWlist.csv}} encapsulate the main parameters of all event from LVK. All these file are extracted from GWOSC the official LVK last realease.
    \item \texttt{\textbf{ORFs}} gather all files related to the overlap reduction functions (ORFs) and its computation. The needed ORFs are all gather in the file \texttt{\textbf{ORF.dat}} where each column correspond to a pair of detector configuration (location and orientation) with the following abrevaitions: H: LIGO Hanford, L: LIGO Livingston, V: Virgo, I: LIGO India, K: Kagra, CE1: Cosmic Exporer at the current Hanford place, CE2: Corsmic explorer at the current Livingsone place, E1, E2, E2, stand for the three corner of the ETtriangle placed in Sardegnia. More details about these function can be fond in \cite{Thrane_2013, Allen_1997}.
    \item \texttt{\textbf{PICs}} contains the power integrated curves (PICs) used to assess the sensitivity of a detector or network of detectors to a stochastic gravitational-wave background.More details about them and theirs computations can be found in \cite{Thrane_2013, Allen_1997, Christensen_1992}.
    \item \texttt{\textbf{PSDs}} contains the noise power spectral densities for the different detectors available in the program. A list is available Section \ref{sec:detection}.
    \item \texttt{\textbf{dl\_z\_table\_Planck\_15.txt}} is a file uned to interpolate the relation between the luminosity distance and the redshift. The original file has been computed with astropy\cite{astropy}.
    \item \texttt{\textbf{factor\_table.dat}} is a file containing pre-compiled factor $f_d$ used for the compilation of realistic SNRs.
\end{itemize}

\end{document}